\documentclass[prb,twocolumn,floats,aps,showpacs]{revtex4}
\usepackage{epsfig}

\def\8{\infty}
\def\oh{\frac{1}{2}}

\def\d{\partial}

\def\undertext#1{\vtop{\hbox{#1}\kern 1pt \hrule}}

\def\VEV#1{\left\langle\,#1\,\right\rangle}

\def\dd#1{\frac{d}{d#1}}
\def\dbyd#1#2{\frac{d#1}{d#2}}

\def\pbyp#1#2{\frac{\partial#1}{\partial#2}}

\def\br{\\ \nonumber & &}

\def\be{\begin{equation}}
\def\ee{\end{equation}}
\def\bea{\begin{eqnarray} & &}
\def\eea{\end{eqnarray}}

\def\rf#1{(\ref{#1})}

\def\t{\tilde}

\def\cH{{\cal H}}

\def\E{{\cal E}}
\def\rfs#1{Eq.~(\ref{#1})}

\begin{document}

\title{Bosonic Excitations in Random Media}

\author {V. Gurarie and J.T. Chalker }
\affiliation{Theoretical Physics, Oxford University, 1 Keble Rd,
Oxford OX1 3NP, United Kingdom}

\date{\today}


\begin{abstract}
We consider classical normal modes and non-interacting
bosonic excitations in disordered systems. We emphasise generic
aspects of such problems and parallels with
disordered, non-interacting systems of fermions, and
discuss in particular the relevance for bosonic excitations of
symmetry classes known in the fermionic context. We
also stress important differences between bosonic and fermionic
problems. One of these follows from the fact that
ground state stability of a system requires all
bosonic excitation energy levels to be positive,
while stability in systems of non-interacting fermions
is ensured  by the exclusion principle,
whatever the single-particle energies.
As a consequence, simple models of uncorrelated disorder
are less useful for bosonic systems
than for fermionic ones, and it is generally important to
study the excitation spectrum in conjunction with the
problem of constructing a disorder-dependent ground state:
we show how a mapping to an operator with chiral symmetry
provides a useful tool for doing this.
A second difference involves the distinction for bosonic systems
between excitations which are Goldstone modes and those which are not.
In the case of Goldstone modes we review established results illustrating
the fact that disorder decouples from excitations in the low frequency limit,
above a critical dimension $d_c$, which in different circumstances
takes the values $d_c=2$ and $d_c=0$. For bosonic excitations which
are not Goldstone modes, we argue that an excitation density varying with
frequency as $\rho(\omega) \propto \omega^4$ is a universal feature
in systems with ground states that depend on the disorder realisation.
We illustrate our conclusions with extensive analytical and some numerical
calculations for a variety of models in one dimension.

\end{abstract}


\maketitle

\section{Introduction}

Excitations in condensed matter systems with quenched disorder
have been a subject of intense study during the last several decades.
Historically, it has been fermionic excitations in random systems that
have received most attention. The reason for this lies in part
with the rapid development of experiments and theory
involving mesoscopic conductors, where the effects of disorder
in phase-coherent electron systems have been studied in
great detail.\cite{Imry}

It is, however, also of considerable interest to study
random systems with bosonic excitations, and
there is an extensive literature treating problems of this
type, too. For instance, the propagation of phonons
in glasses and of electromagnetic waves in media with random
refractive index has long been a subject of active
research,\cite{SJS,PS} and trapping of light via scattering from
disorder is a principle on which
random lasers are based.\cite{Lasers} Other examples of
bosonic excitations in random systems include vibrations of pinned elastic
structures such as charge density waves,\cite{GL}
magnons in diluted antiferromagnets and spin
glasses,\cite{Harris,WW,HS,AM,Ginzburg,Wan,Chernyshev}
and quasiparticles in superfluid
liquid helium permeating a porous medium.\cite{Helium}
To some extent, work on these problems
has focussed on specific features of individual examples,
and given less emphasis to a generic aspects than has
been the case for disordered fermionic problems.

In this paper we emphasise just these generic aspects. In
particular, we examine the relationship between universality
classes identified for fermionic problems and models
for bosonic excitations, as well as features that are
specific to bosonic problems. In addition, we develop a
framework for treating bosonic excitations which we use
to calculate the density of states and localization
properties of one dimensional, randomly pinned systems.
Some of our results have been presented in short form
elsewhere.\cite{GC}

Our discussion is organised as follows. In
Sec.\,\ref{symmetries} we review the symmetry classes established
within random matrix theory for disordered fermionic systems.
We recall in Sec.\,\ref{bosonic} the general form for a quadratic bosonic
Hamiltonian and the Bogoliubov transformation required to diagonalise it,
and show how it is useful to introduce an auxilliary problem with
structure similar to that in the chiral symmetry class.
In Sec.\,\ref{examples} we discuss how various examples
of systems with bosonic excitations fit into this general framework.
Here we emphasise the distinction between excitations that are
Goldstone modes and those that are not, and outline an established
argument that leads in the latter case to the result
$\rho(\omega) \propto \omega^4$. Then, as an illustrative example, we consider
the random field $XY$ spin chain, using our formalism as the basis
for a numerical study of the discrete system,
presented in Sec.\,\ref{discrete}, and
giving an analytical treatment of the continuum limit in
Sec.\,\ref{continuum}, recovering in both cases the behavior
$\rho(\omega) \propto \omega^4$.
In Sec.\,\ref{heisenberg} we consider the random field Heisenberg spin chain
and some related but simpler models, which are of interest because
disorder enters them in a more general way than for the $XY$ model.
Finally, we summarise the relevant experimental situation
in Sec.\,\ref{experiment} and end with concluding remarks in
Sec.\,\ref{concluding}

\section {Symmetries and Disorder}
\label{symmetries}

To provide a context for our discussion of bosonic systems,
we begin by setting out the symmetry classes that are
recognized within random matrix theory for fermionic
systems.\cite{Mehta,Efetov}
Models of non-interacting quasiparticles play an important role in
the study of fermionic excitations in disordered conductors,
insulators and superconductors.
It has long been appreciated that the properties of these models
are controlled by
the discrete symmetries of the single particle
Hamiltonians. Originally
three symmetry classes for random Hamiltonians were
identified.\cite{Wigner,Dyson0}
One class consists of random Hamiltonians that have time-reversal
invariance but no Kramers degeneracy. In an appropriate basis
the Hamiltonian $H$ is real, so that
\be
\label{WD1}
H=H^*.
\ee
Random Hamiltonians which obey Eq.~\rf{WD1} appear,
for example, when studying
disordered conductors without applied magnetic field.
A second symmetry class consists of time-reversal
invariant random Hamiltonians for particles with half-integer spin and
hence Kramers degeneracy. In this case the time-reveral operation
includes spin inversion, and invariance requires
\be
\label{WD2}
H=\sigma_2 H^* \sigma_2\,.
\ee
(Here and in the following, $\sigma_i$
for $i=1,2$ or $3$ represent the conventional Pauli matrices,
acting on a subspace identified by the context.)
In order for Eq.~\rf{WD2} to
be different in an essential way from Eq.~\rf{WD1},
spin rotation invariance must be broken. Thus
this case is of relevance for disordered conductors
with spin-orbit coupling. A third symmetry class arises
when the Hamiltonian has no discrete symmetries.
Examples of these symmetry classes are provided by the three
Wigner-Dyson random matrix ensembles.\cite{Wigner,Dyson0}

More recently, it has been recognized \cite{NS,VZ,Gade,AZ}
that there are seven additional classes of disordered
fermionic Hamiltonians. These arise where there exists
a special reference energy (taken to be zero in the following) for the system,
and a symmetry operation which relates eigenstates
in pairs.
Three of these seven are referred to as chiral symmatery
classes.\cite{NS,VZ,Gade}
Hamiltonians for these classes can be put into the form
\be
\label{Z1}
H = \left( \matrix {0 & Q \cr Q^\dagger & 0 } \right)
\ee
where $Q$ is itself a matrix or an operator.
They obey the symmetry condition
\be
\label{Z1c}
H = - \sigma_3 H \sigma_3\,.
\ee
This ensures that energy levels
of such Hamiltonians appear in pairs $\pm E$,
since if $\psi$ is an eigenfunction with
energy $E$, then $\sigma_3 \psi$  is an eigenfunction with
energy $-E$. The symmetry condition of Eq.~\rf{Z1c},
when combined with either Eq.~\rf{WD1}, or
Eq.~\rf{WD2}, or neither, leads altogether to three chiral
symmetry classes. Chiral Hamiltonians appear as tight-binding models
with only off-diagonal
disorder\cite{OE,Eggarter,Ziman,Gade,Brouwer} and in the problem of classical
diffusion in a random medium.\cite{GB,CW}

The remaining four symmetry classes arise in the study of
disordered superconductors with pairing treated in the mean field approximation.
The Hamiltonian for such a problem has the Bogoliubov-de Gennes structure
\be
\label{Z2}
H = \left( \matrix { h & \Delta \cr \Delta^\dagger & -h^T } \right),
\ee
where the kinetic term $h$ is Hermitian,
while the gap function $\Delta$ is antisymmetric.
This structure leads to another defining symmetry condition,
\be
\label{Z2c}
H= -\sigma_1 H^* \sigma_1\,.
\ee
As for chiral Hamiltonians, the symmetry displayed in
\rfs{Z2c} ensures that energy levels of
Bogoliubov-de Gennes Hamiltonians appear in pairs, $\pm E$.
There are four symmetry classes of such Hamiltonians,
according to whether or not the system has time-reversal
and spin-rotation symmetry, making the total count ten.

A consequence of the conditions
specified in Eqns.~\rf{Z1c} and \rf{Z2c} is that
statistical properties of energy levels and, in
spatially extended systems, the associated eigenfunctions,
are quite different in these additional symmetry
classes near zero energy, compared to properties far from
zero energy, or in the
Wigner-Dyson symmetry classes.\cite{AZ}

It is natural to ask whether this classification
can be extended to problems involving non-interacting
bosonic excitations or, equivalently, classical normal
modes.
At first sight, it might seem that quasiparticle statistics are
unimportant in a non-interacting system. In one crucial
respect, however, this is untrue, since stability of a system
requires bosonic excitation energies to be positive,
while for fermionic excitations stability is guaranteed by the Pauli
exclusion principle, whatever the single-particle energy levels.
This has two implications. First, energy zero emerges as a special point in
the spectrum of bosonic systems, as it does
for the additional fermionic symmetry
classes discussed above. And second, one anticipates that
matrix elements of bosonic Hamiltonians for random systems
will have specific correlations, which ensure positivity of the
spectrum. Thus, while the most general form for a
quadratic bosonic Hamiltonian (see \rfs{osc} below)
is superficially similar to the Bogoliubov-de Gennes
Hamiltonian, and while normal mode
frequencies, like the eigenvalues of Eq.~\rf{Z2}, appear in
pairs $\pm \omega$, matrix elements must satisfy constraints
in order that frequencies are real.
Such a requirement is in stark contrast with the
assumptions of statistical independence used in the
construction of random matrix ensembles for fermionic systems.
To stress the significance of this point, imagine a
treatment of a disordered, interacting system which proceeds
in two stages, by first finding the ground state and then
calculating excitation energies within a harmonic approximation. The spirit
of random matrix theory for fermionic systems is to divorce these two
stages and approach the second one phenomenologically, choosing statistically
independent matrix elements. By contrast, for bosonic systems it is clear that the two
stages cannot be completely separated. Indeed, whereas
for random matrix theory universal spectral properties
follow largely from symmetry and are independent
of the details of the matrix element distribution,
we argue here that for bosonic excitations
which are not Goldsone modes it is the
requirements of stability and the ensuing correlations
in the Hamiltonian for excitations that give rise to
universal spectral properties.

\section{Bosonic Hamiltonians}
\label{bosonic}

In this section we discuss the most general form for bosonic Hamiltonians
and summarise the diagonalisation procedure, following standard lines
(see, for example, Ref.\,\onlinecite{Blaizot}).
We also emphasise the distinction between oscillator frequencies
and stiffnesses. Finally, we show for oscillations about a
stable ground state that it is natural to rewrite the
Hamiltonian in terms of a chiral matrix.

\subsection{Stiffnesses and frequencies}
The most general bosonic Hamiltonian can be written in the equivalent
forms
\begin{eqnarray}
\label{osc}
H &=& \oh \sum_{i,j=1}^N
\left[ M_{ij} p_i p_j + K_{ij} q_i q_j + 2 C_{ij} q_i p_j
\right]
\nonumber \\
&\equiv&
\oh \left( \matrix { {\bf p} & {\bf q} } \right)
\left( \matrix { M & C \cr C^T & K }
\right) \left( \matrix{{\bf p} \cr {\bf q}} \right)
\nonumber \\
&\equiv&
\oh \left( \matrix{{\bf a^\dagger} & {\bf a}} \right) \left(
\matrix { \Gamma & \Delta \cr \Delta^\dagger & \Gamma^T} \right) \left(
\matrix {{\bf a} \cr {\bf a^\dagger}} \right)
\end{eqnarray}
Here, $q_i$ and $p_i$ are the coordinates and momenta of the oscillators,
and $a^\dagger_i, a_i= \left( q_i \pm i p_i \right)/\sqrt{2}$ are
bosonic creation and annihilation operators.
The matrices $M$ and $K$ are real and symmetric,
while $C$ is an arbitrary real matrix. Equivalently, $\Gamma$ is hermitian, while
$\Delta$ is symmetric.  Physically, $M$ is the inverse mass matrix of
the oscillators, $K$ is the matrix of spring constants, and
couplings of the type represented by $C$ occur, for example,
in spin systems. It is convenient to
define the $2N \times 2N$ symmetric matrix
\be
\cH = \left( \matrix { M & C \cr C^T & K} \right).
\ee

We are interested in frequencies of oscillators decribed by Eq.~\rf{osc}.
From Hamilton's equations of motion
it follows that
these frequencies are eigenvalues of the
non-Hermitian matrix
\be
\label{freq1}
\cH' = \left( \matrix {- i C^T & -i K \cr i M & i C } \right)
\equiv \sigma_2 \cH\,,
\ee
They are real if the system is stable, in which case
$H$ is bounded from below,
the eigenvalues of $\cH$ are positive, and we can write
$\cH=Q^T Q$ with $Q$ real.
In thse terms, to find frequencies we need to diagonalize the
matrix $\sigma_2 Q^T Q$, but its eigenvalues coincide with those
of another matrix
\begin{equation}
\label{Omegaconstraint} \Omega=Q \sigma_2 Q^T,
\end{equation}
which is explicitly Hermitian
and, moreover, antisymmetric, so that frequencies come in pairs $\pm
\omega_i$, with $i=1 \ldots N$.

If $\Omega$ is interpreted as a random Hamiltonian, then according
to the classification scheme discussed in Sec.~\ref{symmetries}
it belongs to one of the Bogoliubov-de Gennes
classes (more precisely, to class D, see Ref.~\onlinecite{AZ}). While
this is indeed an indication that random oscillators behave in
many ways like random fermionic Hamiltonians from one of the additional
symmetry classes (having frequencies in pairs,
with $\omega=0$ as a special point in the spectrum),
the identification of $\Omega$ as class D operator is not
by itself necessarily helpful since $\Omega$ does not have statistically
independent matrix elements, but rather
is constrained to have the form given in \rfs{Omegaconstraint}.
Instead, we shall see that a link to matrices wih chiral symmetry
proves more useful.

The computation of oscillator frequencies
can equivalently be described as a Bogoliubov
transformation for coordinates and momenta, specified by real
matrices $g$ which obey
\begin{equation}
\label{bog} \sigma_2  = g \sigma_2 g^T
\end{equation}
and tranform $\cH$ to $g \cH g^T$.
Diagonalizing $\cH$ using this transformation,
the Hamiltonian $H$ of Eq.~\rf{osc} takes the form
\begin{equation}
H = \sum_i |\omega_i| a_i^\dagger a_i\,.
\end{equation}

In addition to the frequencies, obtained as eigenvalues
of $\Omega$ or by Bogoliubov transformation
of $\cH$, the eigenvalues of $\cH$
also have physical significance, for example as inverse static
susceptibilities. We refer to them as
{\sl stiffnesses}, denoting them by $\kappa_i$, $i=1,\ldots 2N$.
In general, there is no simple relationship between
stiffnesses and frequencies, but several special cases
provide important exceptions, as follows.
Consider first $\cH$ with $C=0$ and $M=\openone$,
representing oscillations of particles, all with unit mass, connected by springs
with spring constants $K_{ij}$.
In this example, half of all the stiffnesses are equal to $1$,
while the other half are the eigenvalues $\kappa_i$ of the matrix $K$;
the frequencies and stiffnesses are related by
\be \label{squarerelation}
\omega_i = \pm \sqrt{\kappa_i}\,.
\ee
A second special case arises for magnon excitations in weakly
disordered ferromagnets, which have a Hamiltonian of
the form of Eq.~\rf{osc} with $M=K$ and $C=0$. In that case,
the stiffnesses are the eigenvalues of $M$ and $K$, and come in
identical pairs. The frequencies are simply
\begin{equation}
\omega_i= \pm \kappa_i\,.
\end{equation}
A final and important special case occurs when one stiffness, say
$\kappa_1$, is much smaller than all others. In
this regime, an approximate relation exists between the
smallest frequency and the smallest stiffness:
\begin{equation}
\omega_1 \propto \pm \sqrt{\kappa_1}\,.
\label{kap}
\end{equation}

To derive this relation, a more detailed analysis of the structure of $\cH$
and $\Omega$ is required. Since $\cH$ is a real symmetric matrix with positive
eigenvalues, it can in general be represented as
\be
\cH = U \Lambda^2 U^T
\ee
where $U$ is an orthogonal matrix and
$\Lambda_{ij}=\lambda_i \delta_{ij}$ is the diagonal matrix whose
eigenvalues are square roots of the eigenvalues of $\cH$. The
corresponding matrix $\Omega$ can be written as
\begin{equation}
\Omega = \Lambda U^T {\sigma_2} U \Lambda\,.
\end{equation}
Introducing an antisymmetric matrix $A = U^T \sigma_2 U$, we have
\begin{equation}
\label{freqproof} \Omega_{ij}=\lambda_i A_{ij} \lambda_j\,.
\end{equation}
Suppose initially that one of the stiffnesses vanishes, so that $\lambda_1=0$. Then at
least one frequency must vanish as well, since $\det \, \Omega =
\det \cH$. In fact, because $\Omega$ is antisymmetric and its frequencies come
in opposite pairs, two frequencies are zero. The mathematical
mechanism for this is clear from Eq.~\rf{freqproof}. First, since $\Omega_{1
i} = \Omega_{i 1} = 0$ for all $i$, one of the
eigenvalues of $\Omega$ is $\omega_1=0$, with an eigenvector
$\psi^{(1)}_{i} = \delta_{1i}$. Second, all other eigenvalues can be found
by diagonalizing a smaller matrix, $\Omega'_{ij}$ where $2 \le i, j
\le 2N$. But $\Omega'$ is an odd-dimensional antisymmetric matrix,
and therefore has at least one zero eigenvalue, $\omega_2$,
with an associated normalised eigenvector which we write as
$\psi^{(2)}_{i}$, where $\psi^{(2)}_1 = 0$.
Now treat small non-zero $\lambda_1$ using perturbation theory
about this limit, with $\lambda_1\equiv\epsilon$.
The change in $\Omega$ is
\bea \delta \Omega_{ij} =  \epsilon \left( \delta_{1 i}
A_{ij} \lambda_j  + \lambda_i A_{ij} \delta_{1j} \right)(1-\delta_{1i}\delta_{1j})\,.
\eea
Under this perturbation,
the doubly degenerate eigenvalue $\omega=0$ of $\Omega$ splits
into $\pm \omega_1$, determined by diagonalizing
the $2\times 2$ matrix
\begin{equation}
\epsilon \left( \matrix { 0 & \sum_{i=2}^{2N} A_{1i} \lambda_i
\psi^{(2)}_i \cr - \sum_{i=2}^{2N} A_{1 i} \lambda_i \psi^{(2)}_i
& 0 } \right).
\end{equation}
Hence \begin{equation} \omega_1 = \pm \lambda_1 \left|
\sum_{i=2}^{2N} A_{1 i} \lambda_i \psi^{(2)}_i \right| \propto \pm
\sqrt{\kappa_1}
\end{equation}
barring an accidental vanishing of the matrix element
$\sum_{i=2}^{2N} A_{1 i} \lambda_i \psi^{(2)}_i$.

The usefulness of this result lies in the following.
Consider a random system with bosonic
excitations that are localized with a finite localization length
at low frequency, and let the density of stiffnesses be $d(\kappa)$.
We expect that each localization volume can be treated as an independent system,
and that each low frequency excitation will have a frequency much smaller
than that of other excitations in its own localization volume, and will therefore
be associated with a single small stiffness.
Applying Eq.~\rf{kap}, the density of excitation frequencies $\rho(\omega)$
for small $\omega$ is
\be
\label{ffr}
\rho(\omega) = d(\omega^2) \omega\,.
\ee

One interesting check of these conclusions is provided by random
matrix theory. Consider an ensemble of real random
matrices $Q$, with size $N\gg 1$ and
probability distribution $P \propto \exp \left( -
Q^T Q \right)$. Let $\Delta$ be the typical magnitude of the
eigenvalue of $\cH=Q^TQ$ closest to zero.
For $\cH$ generated in this way, one finds $d(\kappa) \propto
\kappa^{-1/2}$ at both $\kappa \ll \Delta$ and $\kappa \gg \Delta$.
Computing the eigenvalue density of $\Omega=Q \sigma_2 Q^T$ using superintegrals,
one finds that\cite{MZ} $\rho(\omega)$ is independent of $\omega$ for
$\omega \ll \Delta$, while $ \rho(\omega) \propto  \omega^{-1/3}$ for $\omega \gg \Delta$.
It is clear that in the
regime $\omega \ll \Delta$, \rfs{ffr} is indeed applicable, while in
the opposite regime it breaks down.

\subsection{Chiral Symmetry}
\label{chiral-symmetry}

To study stiffnesses of bosonic oscillators with $\cH=Q^T Q$, it  is advantageous to introduce
$\t \cH$, an auxiliary matrix in which $Q$ and $Q^T$ enter linearly.
\be
\label{chi}
\t \cH = \left( \matrix { 0 & Q \cr Q^T & 0 } \right)\,.
\ee
Obviously, the eigenvalues $\lambda_i$ of $\t \cH$ are square roots of the stiffnesses $\kappa_i$
of $\cH$, and the matrix $\t \cH$
plays the role of the square root of the original bosonic Hamiltonian $\cH$.
On the other hand, the off-diagonal structure of $\t \cH$ is the defining feature
of the chiral symmetry class, discussed in Sec.~\ref{symmetries}.
While the direct implications of this connection are limited, because
the elements of $Q$ are not independent random variables as would be the
case in a random matrix ensemble,
techniques originally developed for systems in this symmetry class will prove
useful in our treatment of one-dimensional systems, as we describe in
Sec.~\ref{discrete} and Sec.~\ref{continuum}.

To study the frequencies of the oscillators, as opposed to
their stiffnesses, a second auxiliary matrix
\begin{equation}
\label{chiboson} \t \cH' =  \left( \matrix { 0 & Q \cr \sigma_2
Q^T & 0 } \right)
\end{equation}
is helpful. We shall call
matrices with this structure {\it chiral bosonic
matrices}. The eigenvalue
equation for $\t \cH'$ can be written in the form
\be \left( \matrix { 0 & Q \cr
Q^T & 0 } \right) \psi = \left( \matrix {1 & 0 \cr 0 & \sigma_2}
\right) \sqrt{\omega}~\psi\,, \ee
which of course inherits its structure from Hamilton's equations.

The main difficulty in making use of these ideas is that, in practice,
only $\cH$ is known initially and a method must be developed to find $Q$.
Moreover, $Q$ is defined only up to a left multiplication by an
arbitrary orthogonal matrix. For the introduction of $Q$
to be helpful, it will be important that it can be chosen to have a simple form,
with, for example, only short range couplings. We shall show that this is indeed
possible for a variety of problems.


\section{Bosonic Excitations: General Aspects}
\label{examples}

It is a feature of models for disordered fermionic systems from
the chiral and Bogoliubov-de Gennes symmetry classes that their
characteristic behavior appears only close to the reference energy,
identified by the discrete symmetry of the Hamiltonian, while
spectral properties at energies far from this are indistinguishable from
those of the Wigner-Dyson classes. In a similar way,
for bosonic excitations in disordered systems we expect it to be properties at low frequency that
that are of particular interest. Examples of such excitations
can be divided into two categories, according to whether or not they are
Goldstone modes, associated with a broken continuous symmetry.
In this section we summarize for each of these categories some of the
previously established results. We also illustrate the introduction
of a chiral Hamiltonian $\t \cH$ for a one-dimensional phonon model.

\subsection{Goldstone Modes}

Important instances of Goldstone modes in disordered systems include acoustic phonons
in glasses and alloys,\cite{SJS,PS}
 and spin waves in isotropic, dilute ferromagnets and antiferromagnets,
and isotropic spin glasses.\cite{Harris,WW,HS,AM,Ginzburg,Wan,Chernyshev}
In disorder-free versions of these systems, low frequency
excitations have long wavelength and are described by equations of motion that involve
macroscopic properties of the system: density, elastic constants, magnetic susceptibility and
spin stiffness. Disorder introduces local fluctuations in the values of these quantities, but one expects
excitations to couple only to fluctuations averaged over a volume with linear
dimensions set by a wavelength. Because of this, randomness only weakly affects
low-frequency Goldstone modes, especially in higher dimensions
for which the averaging is most effective. Such averaging is demonstrated by the low-frequency behavior of the
excitation density $\rho(\omega)$, which above a critical dimension $d_c$ varies with the same
power of $\omega$ as in the disorder-free system.
It is also shown by localization properties of excitations: in one and two-dimensional systems,
the localization length $\xi(\omega)$ diverges as $\omega$ approaches zero,
while in higher-dimensional systems, all low-frequency states are extended.
In this subsection we review the behavior of $\rho(\omega)$ and $\xi(\omega)$
for phonons in alloys and for spin waves in diluted antiferromagnets and spin glasses, considering
the effect of weak disorder included in the relevant equation of motion.
We also use the Hamiltonian for phonons in a one-dimensional disordered system as an illustration of
the general mapping to chiral models, introduced in Sec.~\ref{chiral-symmetry}.

Consider first a scalar version of a model for acoustic phonons, in which a mode
with frequency $\omega$ has coordinate $q({\bf r})$ satisfying
\be\label{phonons}
\omega^2 q({\bf r}) = - c^2 \nabla^2 q({\bf r})\,.
\ee
Suppose that the speed of sound, $c$ has random fluctuations in space,
about an average value $c_0$, with only short-range correlations.
Work on this and related problems is reviewed in Ref.~\onlinecite{PS},
and an early treatment of localization in this context was given in Ref.~\onlinecite{SJS}.
The essentials for our purposes are as follows. First, the fluctuations
in $c$, averaged over a $d$-dimensional volume of size set by the wavelength in
the disorder-free system, decrease compared to $c_0$ with
decreasing frequency as $\omega^{d/2}$. Thus in this case the critical dimension
is $d_c=0$, and for any $d>d_c$ the excitation density approaches the form found
without disorder at low frequency. Moreover, from a calculation using the
Born approximation, the Rayleigh scattering rate $\tau^{-1}$ varies as $\tau^{-1} \propto \omega^{d+1}$.
(This dependence combines a factor of $\omega^2$ from the frequency-dependence of coupling to disorder, and a factor of
$\omega^{d-1}$ from the density of final states for scattering processes).
In one dimension, disorder localizes with a localization length proportional to the mean free
path, which here is $c_0\tau$, and so\cite{SJS}
\be
\xi(\omega) \propto \omega^{-2}\,.
\ee
In two dimensions it is a familiar feature of electronic
systems that the localization length is exponentially large in
$k_{\rm F}l$ (where $k_{\rm F}$ is the Fermi wavevector and $l$ is the
mean free path): the equivalent parameter for the phonon problem is
$\omega \tau$, so that\cite{SJS}
\be
\xi(\omega) \propto \exp([\omega_0/\omega]^2)\,,
\ee
where $\omega_0$ is a disorder-strength dependent constant.

These results contrast interestingly with those for an antiferromagnet
which has randomness generated either by site dilution (taken small enough that the system is above the percolation
threshold) or by substitution of impurity spins with
a magnetic moment different from that of the host spins.\cite{Harris,Wan}
In a discussion of the random antiferromagnet, it is useful to begin from the dispersion relation for spin waves in
a two-sublattice {\it ferrimagnet} without disorder. At small wavevector $k$, this has the form
\be\label{ferrimagnet}
\omega^2 + (M_a-M_b)\omega = c^2 k^2\,,
\ee
where $M_a - M_b$ is proportional to the difference between the two sublattice magnetic moments,
which are taken to be oppositely aligned. Setting $M_a=M_b$, one recovers the usual dispersion
relation in an antiferromagnet, with spin-wave speed $c$.
For the random antiferromagnet, independent disorder on the
two sublattices generates random fluctuations in the local value of $M_a-M_b$ about a mean value
of zero. Averaging these fluctuations over a volume of size set by the wavelength in the undiluted
system gives a random variable with an amplitude that scales as $\omega^{d/2}$.
Because of this, disorder appears in \rfs{ferrimagnet} via the term $(M_a-M_b)\omega$, which scales as
$\omega^{d/2+1}$. Above the critical dimension $d_c=2$, disorder is irrelevant in the sense that
this term may be be neglected for small $\omega$ compared to
$\omega^2$.
The value of the critical dimension is also apparent from a Born approximation calculation of the
rate for scattering of spin waves by disorder,\cite{Wan} which yields
$\tau^{-1} \propto \omega^{d-1}$.
Because of this, spin waves have a well-defined wavevector
in the low frequency limit for $d>d_c$, since
$\omega \tau \to \infty$ as $\omega \to 0$.
Applying this approach below the critical dimension, for $d=1$ we see that \rfs{ferrimagnet}
determines a relation between the lengthscale of an excitation, which we denote by $1/k$, and its frequency:
\be
k^{d/2}\omega \propto k^2
\ee
at small $\omega$. This implies
\be\label{rho}
\rho(\omega) \propto \omega^{-1/3}
\ee
and
\be\label{xi}
\xi(\omega) \propto  \omega^{-2/3}\,.
\ee

We expect all these results for low-frequency behavior to be characteristic not only
of disordered antiferromagnets, but also of spin glasses, since the two systems have in
common the crucial feature of a magnetization density that is locally non-zero and random,
but has zero average. A detailed treatment of excitations in spin glasses, however, is
much more difficult than in weakly disordered antiferromagnets, because for spin glasses
the ground state is generally disorder-dependent and unknown. Instead, the established approach is
a hydrodynamic one,\cite{HS,Ginzburg} which leads to linearly dispersing modes with a speed
determined by the macroscopic spin-stiffness and susceptibility.
In the light of our discussion, we expect these hydrodynamic results to be correct for $d>d_c$.
By contrast, for one-dimensional systems, microscopic calculations are possible since
frustration is absent and ground states can be determined. Results from computations\cite{Stinch}
in one-dimensional models
of $\rho(\omega)$ and $\xi(\omega)$ coincide with Eqns.~\rf{rho} and \rf{xi}, above.
Similar calculations are also possible in higher dimensions for the Mattis model,
which shares with one-dimensional models the features that frustration is absent and
the ground state is known for each disorder configuration. In $d=2$ these yield $\rho(\omega) \propto \omega|\log(\omega)|$,
where the logarithm is characteristic of behavior at a critical dimension, and in $d=3$ they give
$\rho(\omega) \propto \omega^2$, in agreement with hydrodynamic theory.\cite{Sher}
A discussion of excitations in the Mattis model, building on the methods described in this paper,
will be presented elsewhere.\cite{AG}

Spin waves have also been investigated\cite{BM} in the infinite-range Heisenberg spin glass:
one of the main findings is a density of states that varies with frequency as
$ \rho(\omega) \propto \omega^{3 \over 2}$. We have not been able to make contact between this result
and the approaches described here.

As a next step, it is interesting to return to acoustic phonons in
disordered systems and use the one-dimensional version of this problem
to illustrate some of the methods set out in Sec.~\ref{bosonic}. In fact, in this context
the mapping to a chiral problem was exploited in celebrated early work
by Dyson,\cite{Dyson} and also in calculations by Ziman.\cite{Ziman}
Consider a one dimensional chain of particles with masses $m_i$, connected by nearest-neighbor springs with spring
constants $k_i$, where $m_i$ and $k_i$ are random and positive.
Let $p_i$ and $q_i$ be the momentum and displacement of the $i$-th particle
The Hamiltonian is
\be \label{Dyson1}
H = \sum_i
\left[ {p^2_i \over 2m_i} + {k_i \over 2} { \left( q_i-q_{i-1}
\right)^2 } \right]\,,
\ee
It is convenient to make the canonical transformation
$p_i \rightarrow \sqrt{m_i} p_i$ and $q_i
\rightarrow q_i/\sqrt{m_i}$, giving
\begin{equation}
\label{Dyson2} H = \sum_i \left[ {p^2_i \over 2} + {k_i \over 2} {
\left( {q_i \over \sqrt{m_i}} -{q_{i-1} \over \sqrt{m_{i-1}} }
\right)^2  } \right]\,,
\end{equation}
which is a particular case of the general
bosonic Hamiltonian \rfs{osc}, in which $M=\openone$ and $C=0$,
so that eigenfrequencies are related to stiffnesses $\kappa$ that are eigenvalues of the
matrix $K$, as in \rfs{squarerelation}.
Further discussion is easier in the continuum limit: replacing the index $i$ with a
continuous coordinate $x$, the Hamiltonian becomes
\be \label{phonon-chain}
H = \int dx~\left[
{p^2(x) \over 2} + {k(x) \over 2} \left(\dd{x} \left[ {q(x) \over
\sqrt{m(x)}} \right] \right)^2 \right]\,.
\ee
To find the stiffnesses, one must solve the eigenvalue equation $Kq(x)=\kappa q(x)$ with
\be \label{ex1}
Kq(x) = -{1 \over \sqrt{m(x)}} \dd{x} k(x) \dd{x} {q(x) \over\sqrt{m(x)}}\,.
\ee
Note that the operator $K$ must be positive definite, in order that the chain is stable.
And indeed, defining $a(x) = \sqrt{k(x)}$ and $b(x)=1/\sqrt{m(x)}$, it may be
expressed as a square, in the form $K=Q^T Q$ with $Q= a(x) (d/d{x}) b(x)$.
Introducing a chiral Hamiltonian, as in \rfs{chi}, we then have
\be\label{ex2}
\left( \matrix{ 0 & a(x) \dd{x} b(x) \cr -b(x) \dd{x}
a(x) & 0} \right) \psi = \omega \psi\,.
\ee

At the equivalent point in his treatment, Dyson distinguishes
between different possible choices for the form of disorder.
In one case, termed {\it type I}, \rfs{ex2} is effectively replaced by
\rfs{chi2} below (with $\langle V(x) \rangle = 0$), leading to the singularity
of \rfs{dyson} in $\rho(\omega)$. An alternative, termed {\it type II},
retains instead \rfs{ex2} with only short-range
correlations in $a(x)$ and $b(x)$, representing disorder that couples only weakly
at small $\kappa$, because it is multiplied by spatial derivatives. This yields\cite{Ziman}
a constant $\rho(\omega)$ at small $\omega$,
in agreement with the general arguments set out following
\rfs{phonons}.
Ziman\cite{Ziman} has given a detailed discussion in this context of
the consequences of different types of disorder.\cite{Comment}

\subsection{Non-Goldstone Low Energy Excitations}
\label{bosonic-non-goldstone}

Without Goldstone modes, the very existence of the low-lying
excitations on which we have focussed our attention is not guaranteed.
In fact, as seems first to have been appreciated in the context of atomic
vibrations in glasses,\cite{IK} disorder itself may provide
a route to a gapless spectrum. The essential
ingredients are that the ground state should depend on the disorder realization,
and that excitations at low frequency should be localized by disorder.
Then it is reasonable to consider excitations within each localization
volume separately, and to expect disorder configurations that support
low frequency excitations to occur with finite probability.
Roughly speaking, these excitations occur in regions where the ground state
configuration is unusally sensitive to small changes in the disorder.
In this section we summarise an established approach to this phenomenon
of disorder-generated low frequency excitations, which concentrates on
a single coordinate and its conjugate momentum.
In subsequent sections we apply the formalism developed in this paper to study
the phenomenon more generally.

Following Il'in and Karpov,\cite{IK} consider a one-dimensional anharmonic oscillator
with momentum $p$, mass $m$, coordinate $q$ and potential $U(q)$.
The Hamiltonian is
\begin{equation}
\label{particle} H = {p^2 \over 2 m} + U(q)\,.
\end{equation}
Choosing $U(q)$ to be a smooth random function, we wish
to study the frequency of small-amplitude oscillations about the absolute minimum
in $U(q)$ and, specifically, the probability distribution of this frequency.
To give a more precise meaning to the notion of a smooth random function, we expand it in
Taylor series
\begin{equation}
\label{taylor1} U(q) = \sum_{n=1}^\infty a_n {q^n \over n!}\,,
\end{equation}
where, to fix the zero of energy,  we set $U(0)=0$.
We take the $a_n$ to be random with joint probability distribution $P(a_1, a_2, \dots)$.
We shall assume that $P(a_1,a_2,\ldots)$ is free of zeros and divergences, but its
detailed form will not be important.

Suppose that $U(q)$ has a minimum at $q=q_0$.
In order to discuss excitations of the oscillator,
we first Taylor expand $U(q)$ around $q=q_0$, writing
\begin{equation}
\label{taylor2} U(q) = \sum_{n=2}^\infty b_n \left( {(q-q_0)^n
\over n!}-{(-q_0)^n \over n!}\right)\,.
\end{equation}
The probability distribution of the coefficients $b_n$ is related to that for the $a_n$ by
\[
P(q_0,b_2,b_3, \dots) = \left| \det {\partial \left(a_1, a_2,
a_3,\dots \right) \over \partial \left(q_0, b_2, b_3, \dots
\right)} \right| P(a_1, a_2, a_3, \dots )\,.
\]
Evaluating the Jacobian, we find
\be P(q_0,b_2, b_3, \dots) =  |b_2| P(a_1, a_2, a_3, \dots)\,. \ee
The probability distribution of $b_2$, the curvature of the potential $U(q)$
at a turning point, is hence
\be \label{prb2} P(b_2) = |b_2|\int dq_0
db_3 db_4 \dots P(a_1,a_2,a_3,\dots)\,. \ee
Under the assumption that $P(a_1,a_2,\ldots)$ is free of zeros and divergences,
this integral remains finite as $b_2 \to 0$, and so for small $b_2$
we have $P(b_2) \propto b_2$.

Small amplitude oscillations about $q_0$ have a frequency
$\omega \propto \sqrt{b_2}$, and it then follows
that the probability distribution of these oscillation frequencies varies for
small $\omega$ as
\begin{equation}
\label{dens01} \rho(\omega) \propto \omega^3\,.
\end{equation}

It is a further restriction to demand that a minimum at $q_0$ is the
{\it global} minimum of $U(q)$. A full treatment of this constraint
would be difficult but is fortunately not necessary: the crucial
condition is that $U(q)$ should have no nearby minima deeper than
the one at $q_0$. For that it is sufficient to truncate the expansion
\begin{eqnarray}
\label{taylor3} U(q) = &U(q_0)+ {b_2 \over 2} (q-q_0)^2+ {b_3 \over 6}
(q-q_0)^3\nonumber \\& + {b_4 \over 24} (q-q_0)^4 + \dots
\end{eqnarray}
at ${\cal O}(q-q_0)^4$. After truncation, we require
$|b_3| < \sqrt{3 b_2 b_4}$ in order that $U(q_0) \leq U(q)$ for all $q$.
This further suppresses the probability density for small curvatures, giving
\[ 
P(b_2) = b_2 \int dq_0  db_4  \dots \int_{-\sqrt{3 b_2
b_4}}^{\sqrt{3 b_2 b_4}} db_3~
 P(a_1,a_2,\dots) \propto b_2^{3 \over 2}.
\]
In turn, this brings the frequency distribution at small $\omega$ to the form
\be
\label{dens0}
\rho(\omega) \propto \omega^4\,.
\ee
One expects that the result of including higher order terms in Eq.~\rf{taylor3},
and of ensuring that no more distant minima are lower than the one at $q_0$,
will be to change the constant of proportionality but not the power in Eq.~\rf{dens0}.

Clearly, a serious limitation of this discussion it that
it is limited to a system with a single coordinate and conjugate
momentum. As with our closely-related discussion of random matrices,
preceding \rfs{ffr}, we expect the result to apply quite generally,
provided excitations are localized with a finite localization length
at low frequency. In these circumstances, the coordinate $q$ is interpreted
as being the relevant degree of freedom for a low-frequency
excitation within a localization volume.
One of our objectives in Sec.~\ref{discrete} and Sec.~\ref{continuum}
is to provide detailed evidence for the more general
relevance of \rfs{dens0}.

\section {Discrete 1D Random Field XY model}
\label{discrete}

In this section and Sec.\ref{continuum} we study the excitations of the
one-dimensional, classical $XY$ spin chain in a random field.
This model provides a simple but non-trivial example of a system
with bosonic excitations which are not Goldstone modes, and
has been discussed previously in Refs.~\onlinecite{AR} and \onlinecite{Fogler}.
In what follows we apply the general approach described in Sec.~\ref{bosonic}.
First, we set out definitions and write down the
Hamiltonian  $\cH$ for small amplitude excitations. Second, we find
a local $Q$ which satisfies $\cH=Q^T Q$. As a result, we map our problem
onto a well-known one, involving a one-dimensional chiral Hamiltonian.
We next review established results for such one-dimensional chiral Hamiltonians,
which can be in one of two regimes, depending on the details of their
disorder distribution. We use the mapping to obtain $\rho(\omega)$
both numerically, and (in Sec.~\ref{continuum}) analytically.

\subsection{Definitions}

The Hamiltonian for the random field $XY$ spin chain is
\be \label{ham}
H = \oh \sum_{i=1}^N \Pi_i^2 -\sum_i \cos(\phi_i-\phi_{i+1}) - \sum_i h_i \cos(\phi_i-\chi_i)
\ee
Here $\Pi_i$ are momenta conjugate to the spin angles $\phi_i$,
exchange energy is represented by $-\cos(\phi_i - \phi_{i+1})$, and
$h_i$ and $\chi_i$ are the amplitude and phase of a
random field. It is convenient to introduce the notation
$I(\phi)\equiv-\cos(\phi)$ and $h_i(\phi) \equiv - h_i \cos (\phi - \chi_i ) $.

While the kinetic energy is quadratic in the momenta $\Pi_i$, the potential energy
is strongly anharmonic in the coordinates $\phi_i$.
We want to find the ground state spin configuration $\phi^0_i$
and the frequencies of oscillations about that ground state.
The ground state configuration satisfies
\be
\label{minim}\left. \pbyp{H}{\phi_i}\right|_{\phi=\phi^0}
= 0\,.
\ee
Expanding $H$ about $\phi_i^0$ to quadratic order, Eq.~\rf{ham} reduces to
an expression of the general form given in Eq.~\rf{osc},
and specified by the matrices $C$, $M$ and $K$. In this case, $C=0$ and
$M=\openone$. The symmetric matrix of spring constants,
$K_{ij}= \left. {\partial^2 H / \partial \phi_i \partial \phi_j}\right|_{\phi=\phi^0}$
is tridiagonal, with no-zero entries
\be
\label{def} K_{ii} = \partial^2_\phi I (\phi^0_i-\phi^0_{i-1}) +
\partial^2_\phi I (\phi^0_i - \phi^0_{i+1}) + \partial^2_\phi
h_i(\phi_i)
\ee
and
\be
\label{def'}
K_{i, i+1} = K_{i+1,i}= -\partial_{\phi}^2 I (\phi^0_i-\phi^0_{i+1})\,.
\ee
As discussed in Sec.~\ref{bosonic}, with $M$, $C$ and $K$ of this form, the
excitation frequencies of the system are $\omega_i=\pm \sqrt{\kappa_i}$, where
the stiffnesses $\kappa_i$ are the eigenvalues of $K$.
Our tasks, then, are the linked ones of determining $\phi_i^0$ and diagonalizing $K$.

\subsection{Mapping onto a chiral problem}
Since $K_{ij}$ is a real positive matrix, it can be written as the
square of another real matrix $Q$, in the form $K=Q^T Q$. Our strategy is
to find $Q$ and then study the related chiral Hamiltonian
\begin{equation}
\label{chiral} \t \cH = \left( \matrix {0 & Q \cr Q^T & 0}
\right)\,.
\end{equation}
As the frequencies coincide with the square roots of the eigenvalues of $K$, we will
use $\omega$ to denote the eigenvalues of $\t \cH$ in this and the following
section.

Because $K$ is tridiagonal,
the matrix $Q$ can be chosen bidiagonal, with non-zero elements
$Q_{ii} \equiv A_i$ and $ Q_{i,i-1} \equiv -B_{i-1}$
which satisfy
\begin{equation}
\label{square} K_{ii} = A_i^2 + B_{i}^2, \ K_{i,i+1} =
-A_{i+1} B_{i}\,.
\end{equation}
To show this and find $A_i$ and $B_i$, it is helpful to use the idea of a partial
energy, first introduced in this context by Feigelman: \cite{Fei}
$\E_i(\phi_i)$ is defined to be the ground state energy of a subsystem
which consists of sites $j\leq i$, considered as a function of the orientation of the
boundary spin, $\phi_i$. Formally
\begin{equation}
\label{parte} \E_i(\phi_i) = \min_{\phi_j, j<i} \sum_{j<i} \left[
I (\phi_j-\phi_{j+1}) + h_j(\phi_j) \right] + h_i(\phi_i)\,.
\end{equation}
It satisfies the recursion relation
\begin{equation} \label{eqde} \E_i(\phi) = \min_{\psi}\left[ I \left( \phi -
\psi \right) + \E_{i-1}(\psi) \right] + h_i(\phi)\,.
\end{equation}
Now let the value of $\psi$ which results from the minimization in \rfs{eqde}
be $\psi_0(\phi)$, and define $\phi_{i-1}$ as a function of $\phi_i$
by $\phi_{i-1}(\phi_i)\equiv \psi_0(\phi_i)$. With this notation,
the condition that the right side of \rfs{eqde} is at a minimum
takes the form
\begin{equation}
\label{eqla} -\partial_\phi
I(\phi_{i}-\phi_{i-1})+\partial_\phi \E_{i-1}(\phi_{i-1}) =0\,.
\end{equation}
By differentiating Eq.~\rf{eqla} with respect to $\phi_{i}$,
remembering that $\phi_{i-1}$ is a function of $\phi_i$ in the
sense described above, we find
\begin{equation}
\dbyd{\phi_{i-1}}{\phi_i}=\frac{\partial^2_\phi I
(\phi_i-\phi_{i-1})}{ \partial^2_\phi I(\phi_i-\phi_{i-1}) +
\partial^2 \E_{i-1}(\phi_{i-1}) }\,.
\end{equation}
In addition, we differentiate \rf{eqde} twice with respect to
$\phi_{i}$, again remembering that $\phi_{i-1}$ is a function of
$\phi_{i}$, to find
\begin{eqnarray}
\label{landis}
\partial^2_\phi h_i(\phi_i) = \partial^2_\phi \E_i(\phi_i)&& \\ - \partial^2_\phi I
(\phi_i-\phi_{i-1})&& \frac{\partial^2_\phi \E_{i-1}(\phi_{i-1})}
{\partial^2_\phi I(\phi_i-\phi_{i-1}) + \partial^2
\E_{i-1}(\phi_{i-1})}\,\nonumber.
\end{eqnarray}
This allows us to solve Eqns.~\rf{def}, \rf{def'} and \rf{square}, obtaining
\begin{eqnarray}
\label{ch1} A_i^2 = \frac{ \left[
\partial^2_\phi I(\phi^0_i-\phi^0_{i-1}) \right]^2 }
{\partial^2_\phi I(\phi^0_i-\phi^0_{i-1}) + \partial^2_\phi
\E_{i-1} (\phi^0_{i-1})}
\end{eqnarray}
and
\begin{eqnarray}
\label{ch1'}
B_{i}^2 = \partial^2_\phi
I(\phi^0_{i+1}-\phi^0_{i}) + \partial^2_\phi \E_{i}(\phi^0_{i})\,,
\end{eqnarray}
which completes the derivation of $Q$.

Let us examine the chiral Hamiltonian $\t \cH$, \rfs{chiral}.
By rearranging its rows and columns it can
be put into the form
\begin{equation}
\label{hopping}
\t \cH = \left( \matrix {0      & B_1   & 0     &  0  &   0  &\dots & 0 & 0 \cr
              B_1   & 0 & A_2   &   0  & 0   &\dots & 0 & 0 \cr
              0     & A_2   & 0     &   B_2 & 0  & \dots & 0 & 0 \cr
                      0          & 0    & B_2   &   0   & A_3  &\dots & 0 & 0 \cr
                      0          & 0    &  0    &  A_3  & 0    & \dots & 0 & 0 \cr
                      \dots      & \dots & \dots & \dots & \dots & \dots & \dots & \dots \cr
            0       & 0 & 0 & 0 & 0 & \dots & 0 & B_N \cr
            0         & 0   & 0     &  0  & 0 & \dots     & B_{N} & 0 }
\right)
\end{equation}
which is familiar as a one-dimensional tight binding model with only
off-diagonal disorder,\cite{Eggarter,Brouwer} and is also referred to as
a random chiral one-dimensional Hamiltonian.

\subsection{Established results for one-dimensional chiral problems
and implications for $XY$ model}
\label{chiral-properties}

There has been extensive previous work on one dimensional models of the type represented by Eq~\rf{hopping},
with disorder in the $A_i$ and $B_i$ which is uncorrelated
and chosen to have a simple, known distribution. The results serve as a
guide for our calculations, and we summarise them here. Parameterise the matrix elements $A_i$ and $B_i$ as
\be
A_i = 1 + a + \delta A_i, \ B_i = 1 - a + \delta B_i
\ee
and take $\delta A_i$ and $\delta B_i$ to be independent Gaussian random
variables with zero mean and standard deviation $\sigma$. Behavior at small $\omega$
is very different according to whether or not  $a$ is zero.

For $a=0$ and $\omega \ll \sigma$, the density of states has a singularity\cite{Eggarter} at $\omega=0$
of a type first obtained in a related problem by Dyson,
\begin{equation}
\label{dyson} \rho(\omega) \propto {1 \over \omega \left| \log^3 \omega
\right| }\,,
\end{equation}
and the localization length of these states diverges for $\omega \to 0$ as
\be \label{dysonloc} \xi(\omega) = |\log(\omega)|\,. \ee

By contrast, for $a \not =0$ (sometimes referred to as the staggered regime \cite{Brouwer})
the density of states varies as\cite{OE}
\begin{equation}
\label{power} \rho(\omega) \propto \omega^\beta, \ \omega \ll a,
\end{equation}
with a power $\beta$ that depends on $a$ and $\sigma$.
The localization length is finite in the small $\omega$ limit, and varies for small $a$ as
$\xi \propto |a|^{-1}$.

While the $A_i$ and $B_i$ which arise in our treatment of the $XY$ model have specific correlations not
present in the chiral problems studied previously, some comparisons are useful. In particular, considering for
simplicity weak disorder, $h_i \ll 1$, the system arising from the $XY$ model turns out to be
in the staggered regime. To see this, note that for weak disorder, $|\phi^0_i- \phi^0_{i-1}| \ll
1$, so that $\partial^2_\phi I(\phi^0_i-\phi^0_{i-1}) \simeq 1 $,
\begin{equation}
A_i \approx 1 - \oh
\partial^2_\phi \E_{i-1} (\phi^0_{i-1}), \ B_i \approx 1 + \oh
\partial^2_\phi \E_i(\phi^0_i)
\end{equation}
and hence  $a = \langle{\partial^2_\phi \E}\rangle/2$.
We argue in Sec.~\ref{continuum} that  $\langle{\partial^2_\phi\E}\rangle>0$.
We therefore expect low frequency states to be localized with localization length
$\xi \propto \langle{\partial^2_\phi \E}\rangle^{-1}$, and with a power law density,
as in Eq.~\rf{power}. Quite separately, if states are localized, the assumptions that led
to result $\rho(\omega) \propto \omega^4$ in Sec.~\ref{examples} are justified,
and so we expect the exponent $\beta=4$. We postpone further analytical work to
Sec.~\ref{continuum}, and first treat the problem numerically.

\subsection{Numerical study of the one-dimensional random field $XY$ model}

Our numerical procedure involves several steps. First, for a system of length $L$,
we construct the function $\E_i(\phi)$ for each $i$ by iterating \rfs{eqde}
from $i=1$ to $i=L$. Then we determine $\phi^0_i$, iterating from $i=L$ to $i=1$ and using
the fact that for each $\phi^0_i$, $\phi^0_{i-1}$ minimises the right-hand side of \rfs{eqde}.
Knowing the ground state spin configuration, we compute the matrix elements appearing in the
chiral Hamiltonian \rfs{hopping}, using Eqns.~\rf{ch1} and \rf{ch1'}. Finally, we employ the transfer matrix
technique developed specifically for such Hamiltonians in Ref.~\onlinecite{Eggarter}
to find the integrated density of states. For the random
field we choose a uniform distribution of $\left[h_i\cos(\chi_i),h_i\sin(\chi_i)\right]$
over a disc of radius $D$, independently for each $i$. We find that it is sufficient
for each disorder strength $D$ to study a single realization in a system of length $L=10^6$.

\begin{figure}[htbp]
\centerline{\epsfxsize=2.0in \epsfxsize=3.0in \epsfbox{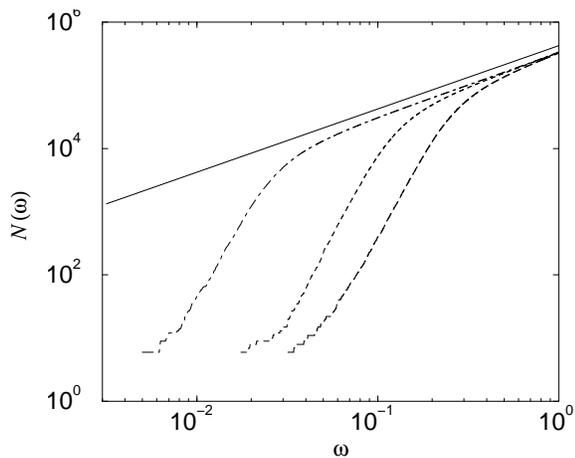}
} \caption{The integrated density $N(\omega)$ plotted as a
function of frequency $\omega$ using logarithmic scales. Dashed,
dotted and dot-dashed lines represent disorder strengths $D= 0.3$,
0.1 and 0.01, respectively. For all three cases, the integrated
density converges at large $\omega$ to that of the disorder-free
system, represented by the full line which has gradient 1.}
\end{figure}

Our results for the integrated density of stiffnesses, $N(\omega) = \int_0^{\omega} d\omega' \rho(\omega')$, \
are shown in Fig 1. Behavior at large $\omega$ approaches that in the disorder-free system, as indicated.
At small $\omega$ we expect the power law
\be
N(\omega) \propto \omega^{{(\beta+1)}}\,.
\ee
This form and also the value $\beta=4$ are both  already apparent in
Fig 1,
as was also the case, at lower precision, in earlier numerical results of Fogler\cite{Fogler} for
much shorter chains.
It is more instructive, however,
to observe that for a power-law density, dimensional analysis fixes the dependence on disorder strength $D$
to be $\rho(\omega) \propto D^{-2 \beta/3} \omega^\beta$.
In Fig. 2 we plot $D^{8/3} N(\omega)$ as a function of $\omega$, showing collapse
of data at small $\omega$ for three different disorder strengths, and power-law
behavior with $\beta=4$.

\begin{figure}[htbp]
\centerline{\epsfxsize=2.0in \epsfxsize=3.0in \epsfbox{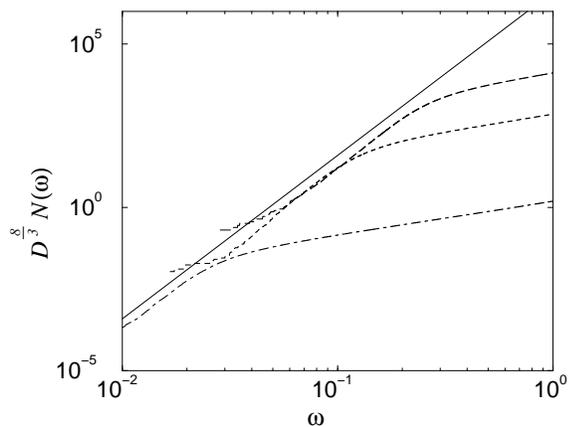}
} \caption{$D^{8 /3} N(\omega)$ vs $\omega$, for disorder strength
$D=0.3$, 0.1 and 0.01, shown with dashed, dotted and dash-dotted
lines, respectively. The straight line has a slope $(\beta+1)=5$.}
\end{figure}

We conclude that, while the exponent $\beta$ appearing in the density of states
is, for a generic chiral problem, disorder-dependent and hence non-universal, the
particular disorder generated in the mapping from the random spin chain, following
Eqns.~\rf{eqde}, \rf{ch1} and \rf{ch1'}, has a specific distribution and the correlations
necessary to produce a universal density of states, $\rho(\omega) \propto \omega^4$.
In the next section we present the analytic derivation of this result.

\section{Analytic treatment of the random field $XY$ model in the continuum limit}
\label{continuum}

In this section we continue our examination of excitations in the one-dimensional random field
$XY$ model, using the continuum limit to make analytic progress. We find
both the localization length as a function of disorder strength and
the density of states as a function of frequency, confirming heuristic arguments
and numerical results given above.

\subsection{Mapping to a chiral problem}
The continuum limit of the random field $XY$ spin chain is reached at
weak disorder, $h_i\ll 1$. In this limit it is
possible to replace the discrete index $i$ with  a continuous variable $x$, and the Hamiltonian
of Eq.~\rf{ham} becomes
\begin{equation}
\label{ham1} H = \int dx \, \left[ \oh \Pi^2 + \oh \left(
\partial_x \phi \right) + h (\phi(x),x) \right]\,,
\end{equation}
with $h(\phi(x),x) = - h(x) \cos \left( \phi(x) - \chi(x) \right)$.
As in the discrete case, we are interested in the
configuration of spins $\phi_0(x)$ that minimizes the potential
energy. This configuration satisfies
\begin{equation}
\label{mimimi} -\partial^2_x \phi_0 (x) + \partial_\phi
h(\phi_0(x),x) =0
\end{equation}
and the amplitudes of normal mode excitations about the
ground-state $\phi_0$ obey
\begin{equation} \label{Schr} \left[ -
\partial^2_x + \partial^2_\phi h(\phi_0(x),x) \right] \psi = \omega^2
\psi\,.
\end{equation}
Just as in the discrete case, because this equation describes deviations
from a minimum, all normal modes have positive stiffness $\kappa = \omega^2$. As a result, the
operator appearing in Eq.~\rf{Schr} can be written as a square. We set
\begin{equation}\label{chi3}
\left[ - \frac{d^2}{dx^2} + \partial^2_\phi h(\phi_0(x),x) \right] =
Q^T Q\,,
\ee
where $Q=-d/dx +V(x)$. In other words, we require a function $V(x)$,
which we term the chiral potential, that satisfies
\be
\label{chidef}
\dbyd{V(x)}{x} + V^2(x) = \partial^2_\phi h(\phi_0(x),x)\,.
\end{equation}

In order to understand properties of the chiral potential,
it is useful to introduce $\E(\phi,x)$, a continuum
version of the partial energy,\cite{Fei} defined following Eq.~\rf{parte} by
\begin{equation}
\label{partec} \E(\varphi,x) = \min_{\phi(x)=\varphi} \int_0^x dy
\,\left[ (\partial_y \phi)^2 + h(\phi(y),y)\right]\,.
\end{equation}
Similarly, the continuum version of Eq.~\rf{eqde} is
\begin{equation}
\label{KPZ}
\partial_x \E + \oh \left( \partial_\phi \E \right)^2 = h(\phi,x)\,,
\end{equation}
which can be thought of as a Hamilton-Jacobi equation for the
action $\E$ of a particle with coordinate $\phi$ moving as a function of time $x$
in the time-dependent potential $h(\phi(x),x)$. In addition, the continuum
version of Eq.~\rf{eqla} relates the ground state configuration to this action, via
\begin{equation}
\label{velo} \dbyd{\phi_0(x)}{x} = \partial_\phi \E (\phi_0(x),x)\,.
\end{equation}
It is easy to check that by writing
\begin{equation}
\label{partialE}
V(x) \equiv \partial^2_\phi \E(\phi_0(x),x)
\end{equation}
we solve Eq.~\rf{chidef}. Thus the chiral potential $V(x)$ may be expressed simply in
terms of the dependence of the ground state energy of a half-system on the boundary spin, $\phi(x)$:
the half-system has coordinate $y$ taking values in the range $0\leq y \leq x$.
With $V(x)$ in hand, we wish to study the continuum
version of Eq.~\rf{hopping}: the chiral operator
\begin{equation}
\label{chi2} \t \cH =\left( \matrix {0 & Q \cr Q^T & 0 } \right) =
\left( \matrix { 0 & -\dd{x}+V(x) \cr \dd{x} +V(x) & 0 }\right)\,,
\end{equation}
which has eigenvalues $\pm \omega$ that are the square roots of those
appearing in Eq.~\rf{Schr}.

\subsection{Treatment of one-dimensional chiral problems}
\label{treatment}

As with the lattice version, discussed in Sec.\ref{chiral-properties},
the spectral properties of $\t \cH$, Eq.~\rf{chi2}, have been studied extensively with
simple choices for the probability distribution of $V(x)$. Behavior is as summarised
for the lattice version in Sec.\ref{chiral-properties}; a particularly detailed study
is given in Ref.~\onlinecite{Comtet}. Here, for completeness we sketch the derivation
of the result that is of most importance for our work: the density of states at low frequency
in the staggered regime, where $\langle V(x)\rangle >0$
for the continuum system plays the same role as
$a>0$ for the lattice model.

Following Ref.~\onlinecite{Comtet}, consider the coupled first-order differential
equations
\begin{equation}
\label{shr1D} \left( \matrix { 0 &- \dd{x} +V(x) \cr  \dd{x} +V(x)
& 0} \right) \left( \matrix{ \psi_1(x) \cr \psi_2(x) } \right) =
\omega \left( \matrix{ \psi_1(x) \cr \psi_2(x) } \right)
\end{equation}
for $x>0$, with boundary conditions $\psi_1(0)=0$, $\psi_2(0)=1$.
From the node counting theorem, the integrated density of states
is given by the density of zeros of $\psi_1(x)$ per unit length.
Introducing the parametrization $\psi_1(x)=\rho(x) \sin
\theta(x), \psi_2(x) = \rho(x) \cos \theta(x)$, one has
\begin{equation}
\label{lang} \dbyd{\theta(x)}{x} = \omega - V(x) \sin \left( 2 \theta(x)
\right)\,.
\end{equation}
Thus we require the average rate of increase in phase with length,  $d\theta(x)/dx$. To find this at small $\omega$,
note that, by assumption, $V(x)$ is mainly positive, and for positive $V(x)$ Eq.~\rf{lang}
has stable fixed points close to $\theta = n \pi$, with $n$ integer.
Rare fluctuations of $V(x)$ which are negative for a long interval in $x$
allow $\theta(x)$ to grow, evolving with increasing $x$ from one such fixed point to the next.
Suppose $\theta(x)$ leaves the vicinity of one fixed point at $x=x_1$
and arrives in the vicinity of the next at $x=x_2$. For $x_1 < x < x_2$
we neglect $\omega$ in Eq.~\rf{lang} and obtain
\begin{equation}
\label{detin} \int { d\theta \over  \sin \left(2 \theta \right)} =
- \int dx~V(x)\,.
\end{equation}
We estimate the integral on the left hand side by noting that the
most important contribution comes from the end points, where
$\theta(x_1) \approx n\pi $, $\theta(x_2) \approx (n+1) \pi $ and
$\sin\left( 2 \theta \right) \sim \omega$. For $\theta$ to increase by $\pi$
we therefore require a fluctuation in $V(x)$ which is sufficiently
negative and extends over a sufficiently large interval in $x$ that
\begin{equation}
\label{fluc} \log \left(\omega\right) > \int_{x_1}^{x_2} dx~
V(x)\,.
\end{equation}
Let $P(\omega)$ be the probability per unit length for such a fluctuation to occur:
the integrated density of states, $\int_0^{\omega}\rho(\omega)d\omega$, is simply $P(\omega)$.
It is natural to expect this probability to be exponentially small in $\log(\omega)$
for small $\omega$, so that, introducing a constant $\alpha$,
\be
\label{P}
P(\omega)\sim \exp(\alpha \log(\omega)) = \omega^\alpha
\ee
and hence
\be
\rho(\omega) \sim \omega^\beta
\ee
with $\beta= \alpha -1$. From an extension of this approach,
one also finds\cite{Comtet} that the localisation length $\xi$
at low frequency varies with the staggering $\langle V(x) \rangle$ as
$\xi \sim \langle V(x) \rangle^{-1}$.


\subsection{Calculation of the density of states}
\label{density-of-states}

A central difficulty of our problem, of course, is that $V(x)$ does not have a simple, given distribution.
Instead, it should be determined by solving \rfs{chidef}, or calculated from the partial energy
using \rfs{partialE}, after in either case first finding the ground
state configuration $\phi_0(x)$, using \rfs{mimimi} or \rfs{velo}.
Our detailed calculations are based on Eqns.~\rf{chidef} and \rf{mimimi}.
Before presenting these calculations, it is useful to develop some qualitative
understanding by an alternative route, using the partial energy, $\E(x,\phi)$,
and its connection to Burgers turbulence.

The evolution of $\E(x,\phi)$ with $x$ is described by Eq.~\rf{KPZ}, which is similar in form
to the much-studied KPZ equation.\cite{KPZ} In this correspondence, $x$ and $\phi$ play the roles of
time and space coordinates, respectively. The stochastic KPZ equation, however, which reads
\begin{equation}
\label{KPZD}
\partial_x \E + \oh \left( \partial_\phi \E \right)^2 - D \partial^2_\phi \E= h(\phi,x)\,,
\end{equation}
differs from \rfs{KPZ} in two respects. One is in the absence of the dissipation term, with coefficient $D$;
the other is in the correlations of the force, $h(\phi,x)$. The absence of
dissipation is of limited importance, because the relationship between the solutions of
\rfs{KPZD} in the small $D$ limit and those of \rfs{KPZ} is well-understood: while \rfs{KPZ}
generally has solutions with many branches, corresponding to local
minima in the energy of the spin chain, by taking the solution of \rfs{KPZD} for $D\to 0$
one finds the envelope of absolute ground states as a function of the boundary spin $\phi(x)$.
For this reason, it is rather natural to study \rfs{KPZD} in our context.

By contrast, the nature of force correlations is more significant: while in standard form
the KPZ equation has a random force $h(\phi,x)$ that is white noise in both $x$ and $\phi$,
our interest lies with correlations that are long-ranged in $\phi$, and have the form
\begin{equation} \label{forcecorr}
\VEV{h(\phi_1,x_1) h(\phi_2,x_2)} = \t h \cos(\phi_1-\phi_2)
\delta(x_1-x_2)\,.
\end{equation}
Such correlations have been studied previously, in the context of Burgers turbulence.
where the Burgers equation, \rfs{Burgers}, (equivalent to \rfs{KPZD} with $D\to 0$) together with
the correlator \rfs{forcecorr}, describes motion of a one-dimensional
fluid.

In this analogy, we view $\phi(x)$ as the coordinate of a particle
as a function of time, $x$, and interpret Eq.~\rf{mimimi} as the equation of motion for the particle.
Imagine a fluid of such particles, without pressure and stirred randomly
with force correlations derived from \rfs{forcecorr}.  Let $u(\phi,x) = d\phi/dx$ be the velocity,
which satisfies the Burgers equation
\begin{equation}
\label{Burgers}
\partial_x u + u \, \partial_\phi u = \partial_\phi h(\phi,x)\,,
\end{equation}
where comparison with Eq.~\rf{KPZ} shows that  $u=\partial_\phi
\E$. The chiral potential is therefore given by $V(x) = \partial_\phi u(\phi_0(x),x)$, the velocity gradient
in a one-dimensional fluid flowing in a time-dependent potential
$h(\phi,x)$, calculated at a point that moves with the
fluid.

From literature on the Burgers equation (see Refs.~\onlinecite{Burgers,Pol,Sinai}),
or alternatively by thinking about the ground state of a spin chain as the boundary
spin $\phi(x)$ is varied,\cite{Fogler} one arrives at the following picture for $\E(\phi,x)$.
As a function of $\phi$ it typically consists of a few piecewise smooth sections,
which meet at cusps that are local maxima of $\E(\phi,x)$. These cusps in $\E(\phi,x)$ are
negative discontinuities
or {\it shocks} of the Burgers velocity field, $u(\phi,x)$, at which
\be
\label{shockwaves}
\lim_{x \rightarrow 0} \left[u(\phi+\epsilon,x) - u(\phi,x)\right] < 0\,.
\ee
They occur at the points where the ground-state spin configuration of the half-chain changes
discontinuously as $\phi$ varies. With increasing system length $x$, they undergo an evolution
in which existing cusps merge and new cusps are born, at matching average rates.
If the trajectory of a particle moving with the Burgers fluid should meet a shock,
the particle remains trapped by the shock for all subsequent $x$. From these statements
it is clear that (for almost all boundary conditions $\phi(L)$) the ground
state trajectory $\phi_0(x)$ does not intersect any shocks. As a next step,
from this we expect that $\langle \partial^2_{\phi}\E(\phi_0(x),x)\rangle > 0$,
on the basis that, first, an average of $\partial^2_{\phi}\E(\phi,x)\equiv \partial_\phi u(\phi,x)$
over $\it all$ $\phi$ must be zero, since $u(\phi,x)$ is periodic in $\phi$, while, second,
an average restricted to $\phi_0$ avoids negative discontinuities of $u(\phi,x)$.
Finally, returning to excitations of the spin chain viewed as a chiral problem,
we conclude that this is in the staggered regime and expect a finite
localization length $\xi \sim \langle
\partial^2_{\phi}\E(\phi_0(x),x)\rangle > 0$;
from dimensional analysis, we expect $\xi \sim {\t h}^{-1/3}$.

Two weaknesses of the argument we have given are clear: first, a
more detailed treatment of averages over ground states $\phi_0(x)$ would be desirable;
and second, it is not certain that behavior known for chiral problems with
disorder uncorrelated in $x$ will necessarily be present in our system,
with correlations of $V(x)$ determined from ground-state properties.
Nevertheless, results of the detailed calculations below are
consistent with the foregoing conclusions.

To make further progress, we need to study statistical properties of $V(x)$.
We find that a direct attack on this problem using the boundary conditions that are physically
appropriate (in which the values of $\phi_0(x)$ are specified at $x=0$ and $x=L$)
is too difficult mathematically. Instead, we approach it indirectly, by relating it
to a version of the problem with simpler boundary
conditions, in which $\phi_0(x)$ and $\partial_x\phi_0(x)$  are specified at $x=0$.
With the latter boundary conditions, we solve jointly \rfs{mimimi} for $\phi_0(x)$
and \rfs{chidef} for $V(x)$. To discuss the connection
between the two alternative sets of boundary conditions,
consider \rfs{chidef} with an arbitrary forcing term, written as $f(x)$:
\be \label{langev}
\dbyd{V(x)}{x} +
V^2(x) = f(x)\,.
\ee
This equation should be integrated from $x=0$ towards $x=L$, with initial condition $V(x)\to \infty$ for
$x \to 0$ since this is the behavior of $\partial_\phi^2\E(\phi_0(x),x)$ at small $x$
and $V(x)$ is related to the partial energy by \rfs{partialE}. Crucially, \rfs{langev}
is unstable in the sense that, if $V(x)$ reaches sufficiently large negative values
for $V^2(x)$ to dominate over $f(x)$, the solution escapes to towards $V(x) = - \infty$,
with the form $V(x) \propto (x-x_0)^{-1}$ as $x$ approaches $x_0$, the position of the instability, from below.
If such an instability is reached, it signals the fact that the trajectory $\phi_0(x)$
no longer represents the absolute ground state of the spin chain but is instead a local maximum
in the energy as a functional of configuration. If the forcing term is derived from the
ground state configuration using $f(x)=\partial^2_{\phi}h(\phi_0(x),x)$, it will have correlations
that ensure this instability is never reached. But if we treat \rfs{langev} as
a Markov process in the way we set out below, some realizations will prove unstable
in this sense. Such trajectories should be discarded, and this can be arranged by
supplementing \rfs{langev} with absorbing boundary conditions at $V(x) = - \infty$.
The surviving trajectories must be weighted in order to sample ground states appropriately,
and we return to this aspect in due course.

First we check that, specifying $\phi_0(x)$ and $\partial_x\phi_0(x)$ at $x=0$
and integrating Eqns.~\rf{mimimi} and \rf{chidef} together, we indeed have a Markov process.
This is seen most clearly by returning to the discrete equations, taking
\rfs{landis} in place of \rfs{chidef} and setting  $\partial^2_\phi I \approx 1$
to obtain
\begin{equation}
V_i = {V_{i-1} \over 1+V_{i-1}} + \d^2_{\phi}h_i (\phi_i)\,.
\end{equation}
Similarly, in place of \rfs{velo}, \rfs{eqla} can be written as
\begin{equation}
\phi_i = \phi_{i-1} + \d_\phi \E_{i-1}(\phi_{i-1})\,.
\end{equation}
It is now clear that, since $\phi_i$ is determined by $h_j$ for $j<i$, it is independent of
the function $h_i(\phi)$. Moreover, by construction, each $h_j(\phi)$ is an independent
random function. Returning to the continuum limit and noting the correlator for
$h(\phi,x)$ given in \rfs{forcecorr}, we see that in \rfs{langev} we should take
$f(x)$ Gaussian distributed, with zero mean and correlator
\begin{equation}
\langle {f(x) f(y)} \rangle = \t h \delta(x-y).
\end{equation}

In light of the discussion given in Sec.~\ref{treatment}, our next step is to
find the probability of an unusually large negative fluctuation in $V(x)$, integrated
over an interval of length $l=x_2 - x_1$.
To this end, it is helpful first to define $S(\alpha)$ by the equation
\be \label{defi}
\VEV{ \exp \left( -\alpha \int_{x_1}^{x_2} dx~V(x) \right)} = \exp \left(
- l S(\alpha) \right)
\ee
where the angular brackets indicate averaging over $V$, and we will find that $S(\alpha)$ is
independent of $l$ when $l$ is large.
We denote this probability (which appeared previously in \rfs{P}) by $P_0(\omega)$ below,
and calculate it using
\be
P_0(\omega) = \max_{l} \int_{-i \infty}^{i\infty} d\alpha\,
\exp\left(-lS(\alpha) + \alpha \log(\omega) \right)\,.
\ee
At small $\omega$ we can approximate the integral by its value at the saddle-point, determined from
\be
-l \pbyp{S}{\alpha} + \log(\omega) =0\,.
\ee
In turn, the maximization on $l$
(remembering that the value of $\alpha$ at the saddle-point is itself a function of $l$) gives
\begin{equation}\label{sadeq}
S(\alpha) = 0.
\end{equation}
Therefore
\begin{equation} P_0(\omega) \propto  \omega^{\alpha_0},
\end{equation}
where $\alpha_0$ is the solution to the Eq.~\rf{sadeq}.

Now we must compute $S(\alpha)$.
A convenient method is to derive from the Langevin equation \rfs{langev} a Fokker-Planck equation,
make a similarity transformation of the latter to obtain a Schr\"odinger equation,
and express this as a path integral. The absorbing boundary
condition at $V=-\infty$ is built in automatically in this approach.
These methods are described, for example in Ref.~\onlinecite{ZinnJustin}.
In this way we find
\begin{widetext}
\begin{equation}
\label{msr} \exp\left( - l S(\alpha) \right) = { \int {\cal D} V
\exp \left[ - \int_0^l dx~\left\{ {1 \over 2 \t h} \left(
\left(\dbyd{V}{x} \right)^2 + V^4  \right) - V \left(1 - \alpha
\right) \right\} \right] \over \int {\cal D} V \exp \left[ -
\int_0^l dx~\left\{ {1 \over 2 \t h} \left( \left(\dbyd{V}{x}
\right)^2 + V^4 \right) - V \right\} \right] }.
\end{equation}
\end{widetext}
The path integrals in this expression are propagators in imaginary
time $x$ for a particle moving with coordinate $V(x)$ in a polynomial potential, which is $V^4 -V(1-\alpha)$
in the case of the numerator and $V^4-V$ in the case of the denominator. For large $l$, both path integrals are
dominated by ground-state contributions, justifying the $l$-dependence displayed in \rfs{defi}.
In this limit, moreover, $S(\alpha)$ is given by the difference in ground-state energies
for the two potentials. It is clear  that $S(\alpha)=S(2-\alpha)$ and that $S(0)=0$.
Hence $\alpha_0=2$ and
\begin{equation}
P_0(\omega) \propto \omega^2.
\end{equation}
It is incidentally also apparent that $\langle V \rangle >0$, confirming our earlier argument that
the chiral description of spin-chain excitations is in the staggered regime.

To complete our calculation of the density of excitations in frequency, a further step
is necessary. We have so far considered spin configurations $\phi_0(x)$ that are
generated in the ensemble of disorder realizations using
fixed values for $\phi_0(x)$ and $\partial_x\phi_0(x)$ at $x=0$, and are local
minima of the energy
but not necessarily the absolute minimum.
We should weight these configurations by a factor $P_1(\omega)$, according to the probability that they appear
in an ensemble with physically appropriate boundary conditions, in which $\phi_0(x)$ is
fixed at $x=0$ and $x=L$. In addition, to obtain properties of excitations about the ground state, we
require a further weighting factor $P_2(\omega)$, involving the probability that a configuration
is the absolute minimum in energy. This is in principle a difficult quantity to determine,\cite{note}
because it involves global features of the system, and we content ourselves
with a heuristic approach which is in a similar spirit to our discussion of
an anharmonic oscillator in Sec.~\ref{bosonic-non-goldstone}. Specifically, we discard all
configurations that have a nearby maximum of the energy. In this way we correctly exclude
local minima that are separated from the ground state by the nearby maximum, but we make errors of two kinds.
First, we fail to exclude local minima that have no nearby maximum
and are separated by a large distance in configuration
space from the true ground state: we assume (as in Sec.~\ref{bosonic-non-goldstone}) that excitations
about such local minima have the same statistical properties as those about the true ground state.
Second, we wrongly exclude true ground states that have nearby maxima: we assume that
those ground states which remain are characteristic of the whole set.
We remark finally that if the factor $P_2(\omega)$ is omitted, we obtain an
excitation density averaged over all configurations that are local energy minima.

To find the $\omega$-dependence of these two weighting factors, we consider a family of
nearby configurations $\phi(s,x)=\phi_0(x)+\eta(s,x)$ for $x_1\le
x \le x_2$, parameterized by $s$. The corresponding family of chiral potentials is
$V(s,x) = V(x) + W(s,x)$, and we restrict attention to small $\eta(s,x)$ and $W(s,x)$.
The weight $P_1(\omega)$ appears because, in disorder realizations which generate negative fluctuations of
$V(x)$, trajectories of $\phi(s,x)$ as a function of $x$ are compressed by an amount which
increases with the size of the integrated potential fluctuation. It is therefore determined by comparing
$\eta(s,x_2)$ with $\eta(s,x_1)$. The weight $P_2(\omega)$ is determined by finding the
probability for escape of $W(s,x)$ to negative infinity, signaling the occurrence of an energy maximum.

We find the evolution of $\eta(s,x)$ and $W(s,x)$ in terms of $V(x)$ by linearizing Eqns. \rf{mimimi} and \rf{chidef}
respectively, obtaining
\begin{equation}
\dbyd{\eta}{x} = V(x) \eta(s,x)
\ee
and
\begin{equation}
\dbyd{W}{x}+ 2 V(x) W(s,x) = \d^3_\phi h(\phi,x) \eta(s,x)\,.
\ee
These equations have the solutions
\begin{equation}
\eta(s,x) = \exp \left[ \int_{x_1}^x dy~V(y) \right] \eta(s,x_1)
\ee
and
\begin{widetext}
\begin{equation}
\label{workW}
W(s,x) = \exp\left({-2 \int_{x_1}^x dy~V(y)}\right) \left( W(s,x_1) +
\eta(s,x_1) \int_{x_1}^{x} dy~
\left[
\exp \left(3 \int_{x_1}^{z} dz~V(z) \right) \d^3_\phi h(\phi_0(y),y) \right] \right).
\ee
\end{widetext}

Let us choose the parametrization $s$ in such a way that at $s=0$,
$\eta(s,x_1)=0$ and $V(s,x_1)$ has a minimum. Then for small $s$,
$\eta(s,x_1) \propto s$ and $W(s,x_1) \propto s^2$. We find
$P_1(\omega)$ as follows. For the integrated value of $V(s,x)$
along a trajectory from the family to have a value similar to that
at $s=0$, the integral of $W(s,x_2)$ must not be too large. We
hence select those values of $s$ for which $\int_{x_1}^{x_2} dx
W(s,x) \lesssim 1$. Using \rfs{workW} and that fact that
$\int_{x_1}^{x_2} dx V(0,x) =\log(\omega)$, this requires $|s|
\lesssim \omega$, which in turn implies that $|\eta(s,x_1)|
\lesssim \omega$ and $|\eta(s,x_2)| \lesssim \omega^2$. Since the
physically relevant boundary conditions for the spin chain fix the
value of $\phi(x)$ at $x=L$ (that is, at larger $x$), we weight
configurations uniformly in $\eta(s,x_2)$ and conclude that
$P_1(\omega) \propto \omega^2$. Turning to $P_2(\omega)$, we
require that $W(s,x_2)$ calculated using our linearization should
not be large and negative for any $s$, since otherwise a full
treatment, including non-linearities, would with high probability
result in escape. Minimizing $W(s,x_2)$ with respect to $s$, using
Eq.~\rf{workW}, we require
\begin{equation}
\min_s W(s,x_2)
\propto - \omega^{-2} I^2 \gtrsim -1\,,
\end{equation}
where
\begin{equation} I=\int_{x_1}^{x_2} dx~\left[ \exp \left( 3
\int_{x_1}^x dy~V(y) \right)
\partial^3_\phi h(\phi_0(x),x) \right]\,.
\end{equation}
For small $\omega$ and large $|x_2-x_1|$, $I$ is a random variable whose distribution
is independent of $\omega$, and hence the
probability that $\min_s W(s,x_2) \gtrsim -1$ is $P_2(\omega) \propto \omega$.

Combining these results with the form derived for $P_0(\omega)$,
we find an integrated density of states for excitations, averaging over all
local minima of \rfs{ham}, which varies as
\begin{equation}
N(\omega) \propto  P_0(\omega) P_1(\omega) = \omega^4\,,
\end{equation}
while for the global minimum we find
\begin{equation}
N(\omega) \propto P_0(\omega) P_1(\omega) P_2(\omega) = \omega^5\,.
\end{equation}
Hence $\beta=3$ or $\beta=4$, in agreement with Eqs.~\rf{dens01}
and \rf{dens0}. Strikingly, the behavior derived here for the random field $XY$ chain
matches that expected from the simple discussion of an anharmonic oscillator,
given in Sec.~\ref{bosonic-non-goldstone}.

\section{Other systems without Goldstone modes}
\label{heisenberg}

In this section we discuss the extent to which the approach we have set out for the
random field $XY$ spin chain can be extended to treat excitations in other models for disordered
systems without a broken continuous symmetry. The $XY$ spin chain has two obvious and important
simplifying features: disorder couples only to one of the dynamical variables ($\phi$ but not $\Pi$),
so that frequencies are simply related to stiffnesses; and the system is one-dimensional.
The calculations we have described involved several steps: mapping to a chiral formulation;
analytic determination of the excitation density in frequency via a study of statistical
properties of disorder within this chiral formulation; and, for an efficient numerical treatment,
the use of a partial energy. As we broaden the range of models under consideration, fewer of these
steps remain possible. We consider, first, a generic one-dimensional continuum system,
with a pair of conjugate dynamical variables at each point and disorder that couples
to both, and second,  a specific example of such a problem, the random field Heisenberg spin chain.
In these cases we find a chiral description and introduce a partial energy, but are not
able to determine statistical properties of the disorder appearing in this description. Instead, we
derive for such problems the general relation between densities of stiffnesses and of excitation
frequencies, suggested from an analysis of random matrix theory in \rfs{ffr} above.
Third, for the random field $XY$ model in two dimensions we show how to introduce a chiral
description, but leave applications of this for future work. For all
these
problems, we expect $\rho(\omega)\propto \omega^4$ at small $\omega$,
on the general grounds discussed in Sec.~\ref{bosonic}.

\subsection{Generic one-dimensional problem}
\label{generic}

Consider a one-dimensional system with conjugate dynamical variables $\phi_1(x)$ and $\phi_2(x)$,
and the Hamiltonian
\be \label{flat}
H = \int dx~\left[ \oh \left\{ \left(\dbyd{\phi_1}{x}\right)^2 + \left(\dbyd{\phi_2}{x}\right)^2 \right\}
 + h (\phi_1, \phi_2,x) \right]\,.
\ee
We wish to study harmonic excitations about the ground state: for $i=1,2$, let
$\phi^0_i$ be the configuration that minimises the energy, \rfs{flat},
and write $\phi_i = \phi^0_i+\psi_i$. Expanding to quadratic order in $\psi_i$,
\be \label{warmup}
\cH = \left( \matrix { -\partial_x^2 +
\d^2_{\phi_1} h & \d_{\phi_1} \d_{\phi_2}h \cr \d_{\phi_1}
\d_{\phi_2}h & - \d_x^2 + \d^2_{\phi_2} h } \right)\,.
\ee
To write this as $\cH = Q^T Q$ and find $Q$, we essentially repeat the sequence of arguments which led
from \rfs{ham1} to \rfs{chi2}. We introduce a partial energy
$\E(\phi_1, \phi_2, x)$ which satisfies
\be \d_x \E + \oh \left\{
\left( \pbyp{\E}{\phi_1} \right)^2 + \left( \pbyp{\E}{\phi_2}
\right)^2 \right\} = h(\phi_1,\phi_2,x)
\ee
and may be used to find the ground state configuration via the analog
of \rfs{velo},
\be
\dbyd{\phi^0_i}{x} = \d_{\phi_i} E(\phi_1^0, \phi_2^0,x)\,.
\ee
Then we define a $2\times 2$ matrix version of the chiral potential, via
\begin{equation}
\label{2chv} V_{ij} =
\partial_{\phi_i} \d_{\phi_j} \E(\phi_1^0, \phi_2^0,x)\,.
\end{equation}
It satisfies
\be \dd{x} V_{ij} + \sum_{k=1}^2 V_{ik} V_{kj} = \d_{\phi_i}
\d_{\phi_j} h(\phi_1,\phi_2,x)
\ee
and therefore $Q$ can be chosen to have elements
\be \label{Q}
\ Q_{ij}  = - \dd{x} \delta_{ij} + V_{ij}\,.
\ee
In this way, we have written \rfs{warmup}
as a type of chiral problem, which we term {\it two-channel}
because $V$ is a $2\times2$ matrix.

Further analysis can be
divided into two stages. One stage, given $V$, is to find
$d(\kappa)$ and $\rho(\omega)$, as was done for the single-channel problem
in Sec.~\ref{treatment}. The other stage
is to determine the distribution
for $V$, as was done for the single-channel problem in
Sec.~\ref{density-of-states}.
General one-dimensional multichannel chiral problems of the type that
arise in our calculation
of stiffnesses have been studied in Refs.~\onlinecite{Brouwer} and
\onlinecite{Brouwer2}, with $V$ chosen Gaussian distributed and
uncorrelated in $x$. However, as far as we are aware, there has been no
previous work on multichannel chiral problems of the type that
give frequencies. For the two-channel problem,
while we have not been able to make
progress in obtaining the distribution of $V$, we have been able to
find a general relation between
calculations of stiffnesses and frequencies. This
connects $d(\kappa)$ and $\rho(\omega)$ in the way given
by \rfs{ffr}.
We present these arguments next.

\subsection{Two-channel Bosonic Chiral Problems}

In this subsection we compare the calculation of
stiffnesses $\kappa\equiv \lambda^2$, determined from the eigenvalue problem
\begin{equation}
\label{2cheqm} \left( \matrix { 0 & -\dd{x} + V \cr \dd{x}+ V^T &
0 } \right) \left( \matrix {\psi_1 \cr \psi_2} \right) = \lambda
\left( \matrix{ \psi_1 \cr \psi_2} \right)\,,
\end{equation}
with the calculation of frequencies $\omega\equiv \epsilon^2$,
determined from
\begin{equation}
\label{2cheqmb} \left( \matrix { 0 & -\dd{x} + V \cr \dd{x}+ V^T &
0 } \right) \left( \matrix{ \psi_1 \cr \psi_2} \right) = \epsilon
\left( \matrix{ \psi_1 \cr \sigma_2 \psi_2 } \right)\,.
\end{equation}
We do this following the technique set out for the single-channel
problem after \rfs{shr1D}, and described for multichannel problems
in Ref.~\onlinecite{Brouwer2}. For both stiffnesses and frequencies,
we write
$\psi_1 = a \psi_2$, where $a$ is an $x$-dependent $2 \times 2$ matrix.
The eigenvalue density, $d(\kappa)$ or $\rho(\omega)$,
is then determined from the evolution of $a$ with $x$,
via a node-counting theorem.
The evolution equation for $a$ is, in the case of stiffnesses, from \rfs{2cheqm},
\begin{equation}
\label{langs} \dbyd{a_s}{x} = \lambda \left( \openone + a_s^2\right) -
a_s V - V^T a_s\,,
\end{equation}
and in the case of frequencies, from \rfs{2cheqmb},
\begin{equation}
\label{langf} \dbyd{a_f}{x} = \epsilon \left( \sigma_2 +
a_f^2\right) - a_f V - V^T a_f\,.
\end{equation}

These equations are the equivalent for the two-channel problem of
\rfs{lang}
for the single-channel problem.
Following Ref.~\onlinecite{Brouwer2}, the density of states is given
by the rate at which the eigenvalues of $a_s$ or $a_f$ move
with increasing $x$ along the real
axis.
Our aim is to compare this rate in the two cases, taking $\kappa$
and $\omega$ small, in a way that does not require detailed knowledge
of the distribution of $V$. We assume only that (as for single-channel
problem) $V$ fluctuates about a non-zero mean,
and that both
the fluctuations and the mean of $V$ have comparable
importance in the evolution of $a_{s}$ and $a_f$ with $x$.

In outline, this evolution is as follows. If
fluctuations in $V$ are omitted, $a_s$ and $a_f$ have stable fixed points
at which both their eigenvalues are small in $\kappa$
and $\omega$ respectively. In both cases, the stable
fixed point has a basin of attraction with a boundary that is reached
when an eigenvalue of $a_s$ or $a_f$ is large and positive
(${\cal O}(\lambda^{-1})$ or ${\cal O}(\epsilon^{-1})$ respectively).
Restoring fluctuations in $V$, with increasing $x$ we find
(partly on the basis of simulations, not presented
here) that $a_s$ or $a_f$ fluctuates in the vicinity of its fixed point for intervals that
are long if $\kappa$ or $\omega$ are small. Each such interval ends
when fluctuations take the matrix to the boundary of the basin of attraction.
One eigenvalue of $a_s$ or $a_f$ then runs off to positive infinity, reappears from
negative infinity and returns to the vicinity of the fixed point.
This
is entirely analogous to the
evolution in the single-channel problem of
$\tan(\theta)$ as $\theta$ increases from $\theta \approx n\pi$ to
$\theta \approx (n+1)\pi$,
which is described in Sec.~\ref{treatment}.

To develop the picture further, we consider separately
the regions close to the fixed points, which are different for
$a_s$ and $a_f$, and the region far from the fixed points,
which is essentially the same in both cases.
It is useful to introduce an explicit
coordinate system. We write
\begin{equation}
V=\oh( \openone +\xi_0 \openone + \xi_1 \sigma_1 + \xi_3 \sigma_3)\,,
\end{equation}
where $\xi_0$, $\xi_1$ and $\xi_3$ are random with mean zero, and
$\langle V \rangle$ is taken proportional to the unit matrix,
with a proportionality
constant that can be changed by a rescaling of the length
coordinate $x$. We also set
\begin{equation}
a_{s,f} = s \openone + z_1 \sigma_1 + z_2 \sigma_2 + z_3 \sigma_3\,.
\end{equation}

For calculation of stiffnesses using this coordinate system,
one has $z_2=0$ and \rfs{langs} becomes
\bea
\dbyd{z_1}{x} = 2 \lambda s z_1 -z_1 + \xi_0 z_1 +\xi_1 s\,, \br
\dbyd{z_3}{x} = 2 \lambda s z_3 - z_3 + \xi_0 z_3 + \xi_3 s\,, \br
\dbyd{s}{x} = \lambda \left( s^2 +
z_1^2 + z_3^2 \right) +\lambda - s + \xi_0 s + \xi_1 z_1+ \xi_3 z_3\,.
\eea
If fluctuations in $V$ are omitted (by setting $\xi_i=0$ for $i=0,1$ and $3$),
these equations have a stable fixed point at $s \approx \lambda$,
$z_1=z_3=0$. Including fluctuations in $V$, the typical magnitudes of
$s$, $z_1$ and $z_3$ are ${\cal O}(\lambda)$.

By contrast, for calculation of frequencies \rfs{langf} gives
\bea \label{a-f}
\dbyd{z_1}{x} = 2 \epsilon s z_1 -  z_1 +\xi_0 z_1 + \xi_1 s\,, \br
\dbyd{z_2}{x} = 2 \epsilon s z_2 + \epsilon-  z_2 + \xi_0 z_2\,, \br
\dbyd{z_3}{x} = 2 \epsilon s z_3 -  z_3 + \xi_0 z_3 + \xi_3 s\,,\br
\dbyd{s}{x} = \epsilon \left( s^2 + z_1^2 + z_2^2 + z_3^2 \right)
-s + \xi_0 s + \xi_1 z_1 + \xi_3 z_3\,.
\eea
In this case, if fluctuations in $V$ are omitted there is a fixed point
at $s\approx \epsilon^3$, $z_2\approx \epsilon$, $z_1=z_3=0$.
Including fluctuations, the typical magnitudes of $s$, $z_1$ and
$z_3$ are ${\cal O}(\epsilon^3)$,
while $z_2$ remains ${\cal O}(\epsilon)$.

Now consider escape of $a_s$ or $a_f$ from the vicinity of the
relevant fixed point, which requires $s \gg \lambda$ in the case
of stiffnesses and $s \gg \epsilon^3$ in the case of frequencies.
In these regimes we argue that the evolution equations in the two
instances are essentially equivalent. More specifically, we make
two approximations. First, we take $z_2=0$ in both cases, even
though this is exact only in the first case. We do so because in
the second case it is clear from \rfs{a-f} that $z_2 \sim
\epsilon$ provided $s<(2\epsilon)^{-1}$, and hence that non-zero
$z_2$ has no important effect on the evolution of $z_1$, $z_3$ and
$s$. Second, for $a_s$ or $a_f$ far from its fixed point, we omit
the terms independent of $a_{s}$ or $a_f$ (and small in $\kappa$
or $\omega$) from the right-hand sides of Eqns. \rf{langs} and
\rf{langf}. With these approximations, and making the rescalings
$a=\lambda a_s$ and $a=\epsilon a_f$, the two equations both
become \be \label{asymptotics} \dbyd{a}{t} = a^2 - aV - V^Ta\,.
\ee This stochastic process results in a stationary probability
distribution for $a$ if absorbing boundary conditions are imposed
at infinity and probability flux is injected near $a=0$. In the
stationary state there is an eigenvalue flux along the positive
real axis, with a rate that is determined by the probability
density in a region near $a=0$. The size of this region, and
therefore the probability density within it, follows in each case
from our discussion of the fixed points and their neighborhoods.
Characterizing $a$ by the value of its trace, and combining the
fixed-point coordinates with the rescalings used to arrive at
\rfs{asymptotics}, the value of ${\rm Tr}\, a$ in this region is
${\cal O}(\lambda^2)$ in the first instance and ${\cal
O}(\epsilon^4)$ in the second. From this, we conclude that the
integrated densities of stiffnesses and frequencies are equal for
$\lambda^2 \sim \epsilon^4$. We hence recover from this discussion
of one-dimensional systems the result given in \rfs{ffr}, which
was reached in Sec.~\ref{bosonic} by a very different route.


For completeness, we remark that the derivation presented here
crucially depends on $\VEV{V(x)} \not = 0$, and assumes that
$\xi_0$, $\xi_1$, and $\xi_3$ become comparable to 1 only during
their rare large fluctuations. And indeed, for $\VEV{V(x)}=0$ it
has been conjectured and checked numerically\cite{GAunp} that in
that case $\rho(\omega) \propto \omega^{-1/3}$ even though it is
well known\cite{Brouwer2} that $d(\kappa) \propto \log(\kappa)$.


\subsection{The random field Heisenberg spin chain}

The random field Heisenberg spin chain is perhaps the most obvious example
of the generic one-dimensional problems discussed above. The conjugate dynamical variables are,
of course, the two components for displacement of a spin from its orientation in the ground state,
and  phase space is made up of spheres for each point on the chain. The curvature of phase space
is responsible for some changes in formuls derived in Sec.~\ref{generic}, which we now set out.

The continuum version of the random field Heisenberg spin chain, parameterizing
spin orientations using the angles $\theta(x)$ and $\phi(x)$, is
\[
\label{Hei}
H = \int dx~\left[\oh \left\{ \left( \dbyd{\theta}{x} \right)^2 + \sin^2\left(\theta \right)
\left( \dbyd{\phi}{x} \right)^2 \right\} + h (\phi, \theta, x) \right]\,.
\]
The evolution equation for the partial energy is
\[
\label{2DKPZ}
\pbyp{\E}{x} + \oh \left\{ \left( \pbyp{\E}{\theta} \right)^2 + {1 \over
\sin^2(\theta)} \left(
\pbyp{\E}{\phi} \right)^2 \right\} = h(\phi,\theta,x)
\]
and, given $\E$, the ground state configuration can be calculated using
\[
\dbyd{\phi_0}{x} = {1 \over \sin^2(\theta)} \d_{\phi} \E, \ \
\dbyd{\theta_0}{x} = \d_\theta \E\,.
\]
Introducing coordinates for deviations from the ground-state configuration,
$\theta = \theta_0 + \psi_\theta$ and $\phi = \phi_0+\psi_\phi$, we take as canonical variables
$\psi_\theta$  and $\psi_\phi \sin \left( \theta_0 \right)$. The quadratic Hamiltonian
can be written $\cH = Q^TQ$, with $Q$ given by \rfs{Q} in terms of the $2 \times 2$ matrix $V$,
which can be computed from $\E$ using
\[
\label{Qhei}
V = \left[ \matrix {  \cot(\theta_0)
+ {1\over \sin^2(\theta_0)} \partial^2_\phi \E &
{1 \over \sin^2(\theta_0) } \partial_\theta
\partial_\phi \E  -2 {\cos(\theta_0) \over \sin^3(\theta_0) } \partial_\phi \E
\cr
{1 \over \sin(\theta_0)}
\partial_\phi \partial_\theta \E &  \partial^2_\theta \E } \right]\,.
\]
This formalism would provide the starting point for a treatment of the
one-dimensional random field Heisenberg model similar to that presented for the
$XY$ model in Sec.~\ref{discrete}.

\subsection{Two-dimensional random field $XY$ model}

The two-dimensional random field $XY$ model provides an example for which
our formulation of excitation problems using chiral Hamiltonians
can be carried through in more than one dimension. The two-dimensional version of \rfs{ham1} is
\[
K = \int  \left\{\oh \left[\Pi^2+ \left( \partial_x \phi \right)^2 + \left( \partial_y \phi \right)^2
\right] + h(\phi,x,y) \right\}dx dy
\]
and the ground state $\phi_0$ satisfies
\be
\label{eqm}
-\left[\frac{\partial^2} {\partial x^2} + \frac{\partial^2} {\partial y^2} \right]
\phi_0 + \partial_\phi h(\phi_0,x,y) = 0\,.
\ee
The amplitudes of normal mode excitations obey
\be
\label{Schr2D}
\left[ - \frac{\partial^2} {\partial x^2}
- \frac{\partial^2} {\partial y^2} + \partial^2_\phi h(\phi,x,y) \right] \psi = E \psi\,.
\ee
We introduce a chiral potential $V(x,y)$, defined as the solution to
\be\label{2dV}
\partial^2_x V + (\partial_x V)^2 + \partial^2_y V + (\partial_y V)^2 = \partial_{\phi}^2 h(\phi_0,x,y)\,,
\ee
and the matrix $Q$, given by
\be
\label{ferm}
Q = \left( \matrix {
-\partial_x + \partial_x V & -\partial_y -\partial_y V \cr
-\partial_y +\partial_y V & \partial_x + \partial_x V } \right)\,.
\ee
Then the matrix $\cH \equiv Q^TQ$ is diagonal, with the operator of \rfs{eqm} as diagonal elements.
The construction of a chiral matrix $\t \cH$ is therefore complete for
this problem, too: $\t \cH$ in fact has the form of a random Dirac
Hamiltonian, studied in Ref.~\onlinecite{LGL}. While
we have not been able to find a quantity for the two-dimensional system that plays the role
of the partial energy in one dimension, we have formulated a generalization of the
Burgers equation, which is sufficient to show that a solution to
\rfs{2dV} exists. The practical determination of the chiral potential,
however, remains an
open problem in this case.

\section{Relation to experiment}
\label{experiment}

The range of physical systems to which the ideas we have set out may
apply is potentially very wide. The essential requirements are:
quenched disorder; a ground state
that can be considered classically;
and excitations that can be treated as weakly interacting normal
modes. These requirements may be met for vibrational modes, either in
glasses, or in randomly pinned phases with broken translational
symmetry such as charge density wave states. They may also
be satisfied for magnetic excitations in disordered ferromagnets
or antiferromagnets, or in spin glasses, either with or
without significant magnetic anisotropy.
The most specific feature for which one would like experimental
evidence is probably the frequency dependence
$\rho(\omega) \propto \omega^4$ of the density for excitations
that are not Goldstone modes.  Alternatively, for excitations
that {\it are} Goldstone modes, interest focusses on
the frequency dependence of the mean free path at small $\omega$,
or, in the case of Heisenberg antiferromagnets and
spin glasses which are quasi one-dimensional,
the low-frequency form of the density of states.

Before summarising the experimental situation,
it is useful to outline ways in which excitations in a disordered
system may fall outside the framework we have used.
Large quantum fluctuations are the obvious route to
different physics, and may be important in at least two ways.
First, it can happen that a ground state is very far from being a classical
configuration dressed with small zero-point fluctuations: random
singlet phases \cite{SH} in disordered quantum spin
chains constitute a well-studied example. Second, it
may be that even low-lying quantized excitations are unlike weakly
interacting bosonic modes. To have a concrete example, consider
the single mode problem for an anharmonic oscillator with potential
$U(q)$, as reviewed in Sec.\,\ref{bosonic-non-goldstone}. Typically, this potential energy
will have local minima separated by barriers from the absolute
minimum, and the importance of quantum
motion through or over the barrier will
depend on the size of the mass $m$ appearing in Eq.\,\rf{particle}.
Thus, for a given $U(q)$, in the semi-classical
limit the low-lying quantum states will
be close to harmonic oscillator levels with the classical
normal mode frequency, while if quantum fluctuations
are large, tunneling through
barriers may hybridise levels associated with different classical
minima, generating two-level systems, or the low-lying levels may be
determined by the form of $U(q)$ in regions too far from its
absolute minimum for the classical normal mode frequency to be
relevant. In the ensemble, this crossover takes place as a function
of frequency, as discussed in Ref.\,\onlinecite{IK}.
As a result, one expects harmonic excitations with a density
$\rho(\omega) \propto \omega^4$
at higher frequencies, and two-level systems
with a constant density at lower frequencies.

Turning to experiments,
somewhat surprisingly, the best evidence that we are aware
of for an excitation density with an $\omega^4$
dependence is from studies of vibrational excitations in
glasses. Here, inelastic neutron scattering
and other measurements \cite{BuchenauPRL}  indicate an excess
density of harmonic modes, compared to what is expected
from Debye theory and the measured speed of sound.
Analysis of the temperature dependence of the heat capacity
\cite{Buchenau} separates two contributions that
are additional to the Debye value. One is
approximately linear in temperature $T$, dominates at low
$T$, and is attributed to two-level systems. The other
varies as $T^5$, dominates at higher $T$, and is attributed to
harmonic modes with the stated frequency dependence for
their density. From a theoretical viewpoint, such
excitations differ in an important way from the one we have discussed,
because they coexist with propagating phonon modes.
The consequences of coupling between the two sets of modes
remain to be investigated, although
coupling between localised {\it anharmonic} vibrational modes
and phonons has been studied in Ref.~\onlinecite{Yu}.

A context in which localised vibrational modes
are expected without any coexisting propagating
phonons is provided by pinned charge density waves,
represented in one dimension by the model studied in Sec.\,\ref{continuum}.
Indeed, it was in this framework that Aleiner and Ruzin \cite{AR}
and Fogler \cite{Fogler} argued for a density of harmonic
proportional to $\omega^4$, with a crossover at low frequency
\cite{Fogler} to a constant density of two-level systems.
As reviewed by Fogler \cite{Fogler},
existing measurements of frequency-dependent
conductivity  \cite{Wu} do not show the response expected from
such harmonic modes, possibly because the low-temperature limit is not
accessed.

Studies of spin waves in disordered magnetic systems
present opportunities to examine both Goldstone
and non-Goldstone modes. In particular, inelastic neutron
scattering measurements of spin dynamics in a dilute
near-Heisenberg antiferromagnet \cite{Uemura} show
the expected broadening in wavevector of excitations
with increasing frequency.
In contrast, magnetic neutron scattering measurements
on amorphous magnetic alloys in which there is
local magnetic anisotropy \cite{Cowley} find
modes which are broad in wavevector at all
frequencies. In this case the density of excitations
is approximately constant in frequency over the measured range.
From our results, we expect a decrease in this density at
low frequency, and
a more detailed examination of low-frequency behavior
would be of considerable interest.

\section{Concluding remarks}
\label{concluding}
Since we have investigated a variety of different
directions in this paper, it is perhaps useful to close
with a short summary of our main points.

Considering the general quadratic Hamiltonian for bosonic
excitations, \rfs{osc}, we have noted a formal similarity between
it and fermionic random matrix Hamiltonians from the additional
symmetry classes (e.g. \rfs{Z2}). We have argued that this has
limited direct consequences, because stability of the ground state
requires correlations between matrix elements of the bosonic
Hamiltonian which invalidate a random matrix approach. Instead we
have shown that there is a useful mapping to an auxiliary problem
with the structure of the chiral symmetry class, which we have set
out explicitly for a range of models.

Examples of bosonic excitations in disordered
media separate into those that are Goldstone modes, and those that
are not. For the former, we
have reviewed established results, which demonstrate that
low frequency excitations decouple from disorder
except for some systems in and below a marginal dimension $d_c=2$.
For excitations that are not Goldstone modes,
we have underlined the way in which disorder itself
generates low frequency excitations, and the universal form
expected for their density, $\rho(\omega) \propto \omega^4$.
Taking as an example the one-dimensional $XY$ model in a
random field, we have used our techniques in detailed analytical and
numerical calculations, obtaining results that illustrate this
behavior of $\rho(\omega)$.

A further application of the techniques we have developed here is
to Mattis models. These model spin glasses lack frustration, and
their statistical mechanical properties are equivalent under a
gauge transformation to those of ferromagnets. They have a ground
state spin configurations which are known for each disorder
configuration, but excitations are nevertheless affected in a
non-trivial way by disorder.\cite{Sher} Because the ground state
is known explicitly, the magnons in Mattis glasses are much easier
to study than in a real spin glass. Being Goldstone modes, these
magnons fall into the same category as those in weakly disordered
antiferromagnets, discussed in this paper. In particular, their
critical dimension is $d_c=2$. In future work \cite{AG} by one of
the present authors and A. Altland, the localization and transport
properties of magnons in two and three dimensional Mattis glasses
will be explored.

\section*{Acknowledgments}

We thank A. Altland and M. Zirnbauer for valuable discussions.
The work was supported by
EPSRC under Grant No. GR/J78327, and in part by the National Science Foundation under
Grant No. PHY99-07949.


\begin{thebibliography}{99}
\bibitem{Comment}
As discussed in Ref.~\onlinecite{Ziman}, a probability distribution
which assigns a large enough value to the probability that $k_i=0$
leads to nonuniversal behavior. This can be understood if one
notices that $k_i=0$ effectively breaks the chain of \rfs{Dyson1}
into noninteracting pieces. It only affects strictly one
dimensional systems and has no equivalent in higher dimensions.
\end{thebibliography}
\end{document}